\documentclass[
onecolumn,
notitlepage,
11pt,
tightenlines,
superscriptaddress,
raggedbottom,
floatfix,
longbibliography,
nofootinbib
]{revtex4-2}

\usepackage{xpatch}

\usepackage{ragged2e}
\usepackage{array}
\usepackage{comment}
\usepackage{soul}

\usepackage{etoolbox}
\patchcmd{\section}{\centering}{\RaggedRight}{}{}
\patchcmd{\subsection}{\centering}{\RaggedRight}{}{}
\patchcmd{\subsubsection}{\centering}{\RaggedRight}{}{}

\usepackage[width=.90\textwidth]{caption}
\usepackage{graphicx} % Required for inserting images
\usepackage{subcaption}
\usepackage[dvipsnames]{xcolor}
\usepackage{amsmath}
\usepackage{amsthm}
\usepackage{enumitem}
\usepackage{amssymb}
\usepackage{physics}
\usepackage[toc,page]{appendix}
\usepackage[
colorlinks,
linkcolor=RedViolet,
citecolor=RedViolet,
bookmarks=false,
urlcolor=Gray,
]{hyperref}

\definecolor{mycyan}{RGB}{35, 250, 221}
\definecolor{myred}{RGB}{255, 79, 79}

\usepackage[framemethod=TikZ]{mdframed}
\newcommand{\be}{\begin{equation}}
\newcommand{\ee}{\end{equation}}

\newcommand{\al}[1]{\begin{align}#1\end{align}}

\DeclareMathOperator{\Obj}{Obj}
\DeclareMathOperator{\Hom}{Hom}

\newcommand{\pd}{{\vphantom{\dagger}}}
\usetikzlibrary{shapes.geometric,calc,backgrounds,intersections,arrows.meta}
\usetikzlibrary{quantikz}
\definecolor{ForestGreen}{rgb}{0.0,0.5,0.0}
\newcommand{\po}{{\phantom{\overline{}}}}
\usepackage{bm}

\newcommand{\deformedlattice}{\scalebox{0.45}{\begin{tikzpicture}[baseline=-0.25cm]
        
     \draw[ultra thick] (-2,0) -- (-1,-1) -- (1,-1) -- (2,0) -- (1,1) -- (-1,1) -- cycle;
    \draw[ultra thick] (-2,0) -- (-3,0);
    \draw[ultra thick] (-1,-1) -- (-2,-2);
    \draw[ultra thick] (1,-1) -- (2,-2);
    \draw[ultra thick] (2,0) -- (3,0);
    \draw[ultra thick] (1,1) -- (2,2);
    \draw[ultra thick] (-1,1) -- (-2,2);

    \draw (0,1.4) node {\huge $a$};
    \draw (-1.75,0.85) node {\huge $b$};
    \draw (-2.1,1.5) node {\huge $g$};
    \draw (-2.5,0.55) node {\huge $h$};
    \draw (-1.85,-0.75) node {\huge $c$};
    \draw (-2,-1.5) node {\huge $i$};
    \draw (0,-1.4) node {\huge $d$};
    \draw (2,-1.4) node {\huge $j$};
    \draw (1.78,-0.68) node {\huge $e$};
    \draw (1.85,0.75) node {\huge $f$};
    \draw (2.5,0.5) node {\huge $k$};
    \draw (2,1.6) node {\huge $l$};
    
   \draw[ultra thick,-{Stealth[length=3mm]}] (0,1) -- (0.1,1);
   \draw[ultra thick,-{Stealth[length=3mm]}] (0,-1) -- (0.1,-1);
   \draw[ultra thick,-{Stealth[length=3mm]}] (2.5,0) -- (2.6,0);
   \draw[ultra thick,-{Stealth[length=3mm]}] (-2.5,0) -- (-2.4,0);
   \draw[ultra thick,-{Stealth[length=3mm]}] (1.5,-0.5) -- (1.6,-0.4);
   \draw[ultra thick,-{Stealth[length=3mm]}]  (-1.5,-0.5) -- (-1.4,-0.6);
   \draw[ultra thick,-{Stealth[length=3mm]}]  (-1.55,-1.55) -- (-1.45,-1.45);
   \draw[ultra thick,-{Stealth[length=3mm]}]  (1.55,-1.55) -- (1.65,-1.65);
   \draw[ultra thick,-{Stealth[length=3mm]}]  (-1.55,1.55) -- (-1.45,1.45);
   \draw[ultra thick,-{Stealth[length=3mm]}]  (1.55,1.55) -- (1.65,1.65);
   \draw[ultra thick,-{Stealth[length=3mm]}]  (1.5,0.5) -- (1.6,0.4);
   \draw[ultra thick,-{Stealth[length=3mm]}]  (-1.5,0.5) -- (-1.4,0.6);

\end{tikzpicture}}}

\newcommand{\deformedlatticeloop}{\scalebox{0.45}{\begin{tikzpicture}[baseline=-0.25cm]
     \draw[thick,orange] (0,0) ellipse (1.5cm and 0.75cm);
     \draw[orange] (-0.75,0.25) node {\huge $\alpha$};
     \draw[ultra thick, orange,-{Stealth[length=3mm]}] (-0.85,0.62) -- (-0.94,0.58);
        
     \draw[ultra thick] (-2,0) -- (-1,-1) -- (1,-1) -- (2,0) -- (1,1) -- (-1,1) -- cycle;
    \draw[ultra thick] (-2,0) -- (-3,0);
    \draw[ultra thick] (-1,-1) -- (-2,-2);
    \draw[ultra thick] (1,-1) -- (2,-2);
    \draw[ultra thick] (2,0) -- (3,0);
    \draw[ultra thick] (1,1) -- (2,2);
    \draw[ultra thick] (-1,1) -- (-2,2);

    \draw (0,1.4) node {\huge $a$};
    \draw (-1.75,0.85) node {\huge $b$};
    \draw (-2.1,1.5) node {\huge $g$};
    \draw (-2.5,0.55) node {\huge $h$};
    \draw (-1.85,-0.75) node {\huge $c$};
    \draw (-2,-1.5) node {\huge $i$};
    \draw (0,-1.4) node {\huge $d$};
    \draw (2,-1.4) node {\huge $j$};
    \draw (1.78,-0.68) node {\huge $e$};
    \draw (1.85,0.75) node {\huge $f$};
    \draw (2.5,0.5) node {\huge $k$};
    \draw (2,1.6) node {\huge $l$};
    
   \draw[ultra thick,-{Stealth[length=3mm]}] (0,1) -- (0.1,1);
   \draw[ultra thick,-{Stealth[length=3mm]}] (0,-1) -- (0.1,-1);
   \draw[ultra thick,-{Stealth[length=3mm]}] (2.5,0) -- (2.6,0);
   \draw[ultra thick,-{Stealth[length=3mm]}] (-2.5,0) -- (-2.4,0);
   \draw[ultra thick,-{Stealth[length=3mm]}] (1.5,-0.5) -- (1.6,-0.4);
   \draw[ultra thick,-{Stealth[length=3mm]}]  (-1.5,-0.5) -- (-1.4,-0.6);
   \draw[ultra thick,-{Stealth[length=3mm]}]  (-1.55,-1.55) -- (-1.45,-1.45);
   \draw[ultra thick,-{Stealth[length=3mm]}]  (1.55,-1.55) -- (1.65,-1.65);
   \draw[ultra thick,-{Stealth[length=3mm]}]  (-1.55,1.55) -- (-1.45,1.45);
   \draw[ultra thick,-{Stealth[length=3mm]}]  (1.55,1.55) -- (1.65,1.65);
   \draw[ultra thick,-{Stealth[length=3mm]}]  (1.5,0.5) -- (1.6,0.4);
   \draw[ultra thick,-{Stealth[length=3mm]}]  (-1.5,0.5) -- (-1.4,0.6);

\end{tikzpicture}}}

\newcommand{\deformedlatticeresolved}{\scalebox{0.45}{\begin{tikzpicture}[baseline=-0.25cm]
        
     \draw[ultra thick, cyan] (-2,0) -- (-1,-1) -- (1,-1) -- (2,0) -- (1,1) -- (-1,1) -- cycle;
    \draw[ultra thick] (-2,0) -- (-3,0);
    \draw[ultra thick] (-1,-1) -- (-2,-2);
    \draw[ultra thick] (1,-1) -- (2,-2);
    \draw[ultra thick] (2,0) -- (3,0);
    \draw[ultra thick] (1,1) -- (2,2);
    \draw[ultra thick] (-1,1) -- (-2,2);

    \draw (0,1.5) node {\huge $a'$};
    \draw (-1.75,0.85) node {\huge $b'$};
    \draw (-2.1,1.5) node {\huge $g$};
    \draw (-2.5,0.55) node {\huge $h$};
    \draw (-1.95,-0.75) node {\huge $c'$};
    \draw (-2,-1.5) node {\huge $i$};
    \draw (0,-1.45) node {\huge $d'$};
    \draw (2,-1.4) node {\huge $j$};
    \draw (1.78,-0.68) node {\huge $e'$};
    \draw (1.9,0.85) node {\huge $f'$};
    \draw (2.5,0.5) node {\huge $k$};
    \draw (2,1.6) node {\huge $l$};
    
   \draw[ultra thick,cyan,-{Stealth[length=3mm]}] (0,1) -- (0.1,1);
   \draw[ultra thick, cyan,-{Stealth[length=3mm]}] (0,-1) -- (0.1,-1);
   \draw[ultra thick,-{Stealth[length=3mm]}] (2.5,0) -- (2.6,0);
   \draw[ultra thick,-{Stealth[length=3mm]}] (-2.5,0) -- (-2.4,0);
   \draw[ultra thick,cyan,-{Stealth[length=3mm]}] (1.5,-0.5) -- (1.6,-0.4);
   \draw[ultra thick,cyan,-{Stealth[length=3mm]}]  (-1.5,-0.5) -- (-1.4,-0.6);
   \draw[ultra thick,-{Stealth[length=3mm]}]  (-1.55,-1.55) -- (-1.45,-1.45);
   \draw[ultra thick,-{Stealth[length=3mm]}]  (1.55,-1.55) -- (1.65,-1.65);
   \draw[ultra thick,-{Stealth[length=3mm]}]  (-1.55,1.55) -- (-1.45,1.45);
   \draw[ultra thick,-{Stealth[length=3mm]}]  (1.55,1.55) -- (1.65,1.65);
   \draw[ultra thick,cyan,-{Stealth[length=3mm]}]  (1.5,0.5) -- (1.6,0.4);
   \draw[ultra thick,cyan,-{Stealth[length=3mm]}]  (-1.5,0.5) -- (-1.4,0.6);

\end{tikzpicture}}}

\newcommand{\fermionloopzero}{\scalebox{1}{\begin{tikzpicture}[baseline=0cm]
     \draw[line width=1pt] (-1,0) -- (0,-0.75) -- (1,0) -- (0,0.75) -- cycle;
    
    \draw[thick] (0,0.75) -- (0,0);
    \draw[thick] (0,-0.75) -- (0,-1.5);
    \draw[ thick] (-1,0) -- (-1,-0.75);
    \draw[ thick] (1,0) -- (1,-0.75);
    
    \draw[line width=1pt] (-1,0) -- (-1.33,0.25);
    \draw[line width=1pt] (-1,0) -- (-1.33,-0.25);
    \draw[line width=1pt] (1,0) -- (1.33,0.25);
    \draw[line width=1pt] (1,0) -- (1.33,-0.25);
    \draw[line width=1pt] (0,0.75) -- (0.25,0.95);
    \draw[line width=1pt] (0,0.75) -- (-0.25,0.95);
    \draw[line width=1pt] (0,-0.75) -- (0.25,-0.95);
    \draw[line width=1pt] (0,-0.75) -- (-0.25,-0.95);

    \draw[thick,orange] (0,0) ellipse (0.6cm and 0.4cm);
    
    \draw[orange] (0.05,-0.55) node {\tiny $\psi$};
    \draw (0,-0.15) node {\tiny $\rho$};
    
     \draw (-0.55,0.55) node {\small $a$};
     \draw (-0.55,-0.55) node {\small $b$};
     \draw (0.6,0.55) node {\small $d$};
     \draw (0.55,-0.55) node {\small $c$};
     \draw (-0.35,1) node {\small $e$};
     \draw (0.35,1) node {\small $f$};

\end{tikzpicture}}}

\newcommand{\fermionloopone}{\scalebox{1}{\begin{tikzpicture}[baseline=0cm]
    \draw[line width=1pt] (-1,0) -- (0,-0.75) -- (1,0) -- (0,0.75) -- cycle;
    
    \draw[thick] (0,0.75) -- (0,0);
    \draw[thick] (0,-0.75) -- (0,-1.5);
    \draw[ thick] (-1,0) -- (-1,-0.75);
    \draw[ thick] (1,0) -- (1,-0.75);
    
    \draw[line width=1pt] (-1,0) -- (-1.33,0.25);
    \draw[line width=1pt] (-1,0) -- (-1.33,-0.25);
    \draw[line width=1pt] (1,0) -- (1.33,0.25);
    \draw[line width=1pt] (1,0) -- (1.33,-0.25);
    \draw[line width=1pt] (0,0.75) -- (0.25,0.95);
    \draw[line width=1pt] (0,0.75) -- (-0.25,0.95);
    \draw[line width=1pt] (0,-0.75) -- (0.25,-0.95);
    \draw[line width=1pt] (0,-0.75) -- (-0.25,-0.95);
    
    \draw[thick, orange] (0,0.6) ..  controls +(220:2.5cm) and +(350:2cm) .. (0,0.3);
    \draw[thick, orange] (0,0.5) ..  controls +(220:1cm) and +(350:0.5cm) .. (0,0.15);
    
    \draw[thick, cyan] (0,0.125) -- (0,0.315);
    \draw[thick, cyan] (0,0.48) -- (0,0.61);
    
    \draw[orange] (0.1,-0.4) node {\tiny $\psi$};
    \draw (0,-0.15) node {\tiny $\rho$};
    
     \draw (-0.55,0.55) node {\small $a$};
     \draw (-0.55,-0.55) node {\small $b$};
     \draw (0.6,0.55) node {\small $d$};
     \draw (0.55,-0.55) node {\small $c$};
     \draw (-0.35,1) node {\small $e$};
     \draw (0.35,1) node {\small $f$};

\end{tikzpicture}}}

\newcommand{\fermionlooptwo}{\scalebox{1}{\begin{tikzpicture}[baseline=0cm]
    \draw[line width=1pt] (-1,0) -- (0,-0.75) -- (1,0) -- (0,0.75) -- cycle;
    
    \draw[thick] (0,0.75) -- (0,0);
    \draw[thick] (0,-0.75) -- (0,-1.5);
    \draw[ thick] (-1,0) -- (-1,-0.75);
    \draw[ thick] (1,0) -- (1,-0.75);
    
    \draw[line width=1pt] (-1,0) -- (-1.33,0.25);
    \draw[line width=1pt] (-1,0) -- (-1.33,-0.25);
    \draw[line width=1pt] (1,0) -- (1.33,0.25);
    \draw[line width=1pt] (1,0) -- (1.33,-0.25);
    \draw[line width=1pt] (0,0.75) -- (0.25,0.95);
    \draw[line width=1pt] (0,0.75) -- (-0.25,0.95);
    \draw[line width=1pt] (0,-0.75) -- (0.25,-0.95);
    \draw[line width=1pt] (0,-0.75) -- (-0.25,-0.95);
    
    \draw[thick, orange] (0,0.6) ..  controls +(220:2.3cm) and +(330:2.2cm) .. (0,0.5);
    \draw[thick, orange] (0,0.3) ..  controls +(220:0.5cm) and +(340:0.3cm) .. (0,0.15);
    
    \draw[thick, cyan] (0,0.125) -- (0,0.315);
    \draw[thick, cyan] (0,0.48) -- (0,0.61);
    
    \draw[orange] (0.05,-0.55) node {\tiny $\psi$};
    \draw (0,-0.15) node {\tiny $\rho$};
    
     \draw (-0.55,0.55) node {\small $a$};
     \draw (-0.55,-0.55) node {\small $b$};
     \draw (0.6,0.55) node {\small $d$};
     \draw (0.55,-0.55) node {\small $c$};
     \draw (-0.35,1) node {\small $e$};
     \draw (0.35,1) node {\small $f$};

\end{tikzpicture}}}

\newcommand{\fermionloopthree}{\scalebox{1}{\begin{tikzpicture}[baseline=0cm]
    \draw[thick] (0,0.75) -- (0,0);
    \draw[thick] (0,-0.75) -- (0,-1.5);
    \draw[ thick] (-1,0) -- (-1,-0.75);
    \draw[ thick] (1,0) -- (1,-0.75);
    
    \draw[line width=1pt] (-1,0) -- (-1.33,0.25);
    \draw[line width=1pt] (-1,0) -- (-1.33,-0.25);
    \draw[line width=1pt] (1,0) -- (1.33,0.25);
    \draw[line width=1pt] (1,0) -- (1.33,-0.25);
    \draw[line width=1pt] (0,0.75) -- (0.25,0.95);
    \draw[line width=1pt] (0,0.75) -- (-0.25,0.95);
    \draw[line width=1pt] (0,-0.75) -- (0.25,-0.95);
    \draw[line width=1pt] (0,-0.75) -- (-0.25,-0.95);
    
    \draw[line width=1pt, cyan] (-1,0) -- (0,-0.75) -- (1,0) -- (0,0.75) -- cycle;
    
    \draw[thick,orange] (0,0.3) ..  controls +(220:0.5cm) and +(340:0.3cm) .. (0,0.15);
    
    \draw[thick, cyan] (0,0.125) -- (0,0.315);
    
    \draw (0,-0.15) node {\tiny $\rho$};
    
     \draw (-0.75,0.55) node {\tiny$a\times \psi$};
     \draw (-0.6,-0.65) node {\tiny $b\times\psi$};
     \draw (0.75,0.55) node {\tiny $d\times\psi$};
     \draw (0.65,-0.6) node {\tiny $c\times\psi$};
     \draw (-0.35,1) node {\tiny $e$};
     \draw (0.35,1) node {\tiny $f$};

\end{tikzpicture}}}

\newcommand{\fermionloopfour}{\scalebox{1}{\begin{tikzpicture}[baseline=0cm]
    \draw[thick] (0,0.75) -- (0,0);
    \draw[thick] (0,-0.75) -- (0,-1.5);
    \draw[ thick] (-1,0) -- (-1,-0.75);
    \draw[ thick] (1,0) -- (1,-0.75);
    
    \draw[line width=1pt] (-1,0) -- (-1.33,0.25);
    \draw[line width=1pt] (-1,0) -- (-1.33,-0.25);
    \draw[line width=1pt] (1,0) -- (1.33,0.25);
    \draw[line width=1pt] (1,0) -- (1.33,-0.25);
    \draw[line width=1pt] (0,0.75) -- (0.25,0.95);
    \draw[line width=1pt] (0,0.75) -- (-0.25,0.95);
    \draw[line width=1pt] (0,-0.75) -- (0.25,-0.95);
    \draw[line width=1pt] (0,-0.75) -- (-0.25,-0.95);
    
    \draw[line width=1pt, cyan] (-1,0) -- (0,-0.75) -- (1,0) -- (0,0.75) -- cycle;

     \draw (-0.75,0.55) node {\tiny$a\times \psi$};
     \draw (-0.6,-0.65) node {\tiny $b\times\psi$};
     \draw (0.75,0.55) node {\tiny $d\times\psi$};
     \draw (0.65,-0.6) node {\tiny $c\times\psi$};
     \draw (-0.35,1) node {\tiny $e$};
     \draw (0.35,1) node {\tiny $f$};

\end{tikzpicture}}}

\newcommand{\fermionloopfive}{\scalebox{1}{\begin{tikzpicture}[baseline=0cm]
    \draw[thick] (0,0.75) -- (0,0);
    \draw[thick] (0,-0.75) -- (0,-1.5);
    \draw[ thick] (-1,0) -- (-1,-0.75);
    \draw[ thick] (1,0) -- (1,-0.75);
    
    \draw[line width=1pt] (-1,0) -- (-1.33,0.25);
    \draw[line width=1pt] (-1,0) -- (-1.33,-0.25);
    \draw[line width=1pt] (1,0) -- (1.33,0.25);
    \draw[line width=1pt] (1,0) -- (1.33,-0.25);
    \draw[line width=1pt] (0,0.75) -- (0.25,0.95);
    \draw[line width=1pt] (0,0.75) -- (-0.25,0.95);
    \draw[line width=1pt] (0,-0.75) -- (0.25,-0.95);
    \draw[line width=1pt] (0,-0.75) -- (-0.25,-0.95);
    
    \draw[line width=1pt] (-1,0) -- (0,-0.75) -- (1,0) -- (0,0.75) -- cycle;
    
     \draw[red] (-0.5,-0.38) node {$X$};
     \draw[red] (0.5,-0.38) node {$X$};
     \draw[ForestGreen] (0.55,0.4) node {$Y$};
     \draw[ForestGreen] (-0.5,0.4) node {$Y$};
     \draw[blue] (-0.25,1) node {$Z$};
     \draw[blue] (0.25,1) node {$Z$};

\end{tikzpicture}}}

\newcommand{\lattice}{\scalebox{0.45}{\begin{tikzpicture}[baseline=-0.25cm]
    \draw[ultra thick] (-2,0) -- (-1,-1) -- (1,-1) -- (2,0) -- (1,1) -- (-1,1) -- cycle;
    \draw[ultra thick] (-2,0) -- (-4,0);
    \draw[ultra thick] (-1,-1) -- (-2,-2);
    \draw[ultra thick] (1,-1) -- (2,-2);
    \draw[ultra thick] (2,0) -- (4,0);
    \draw[ultra thick] (1,1) -- (2,2);
    \draw[ultra thick] (-1,1) -- (-2,2);

    \draw[thick] (-3,0) -- (-3,-1);
    \draw[thick] (0,1) -- (0,0);
    \draw[thick] (0,-1) -- (0,-2);
    \draw[thick] (3,0) -- (3,-1);
    
   \draw[ultra thick,-{Stealth[length=3mm]}] (-0.45,1) -- (-0.35,1);
   \draw[ultra thick,-{Stealth[length=3mm]}] (0.55,1) -- (0.65,1);
   \draw[ultra thick,-{Stealth[length=3mm]}] (-0.45,-1) -- (-0.35,-1);
   \draw[ultra thick,-{Stealth[length=3mm]}] (0.55,-1) -- (0.65,-1);
   \draw[ultra thick,-{Stealth[length=3mm]}] (2.55,0) -- (2.65,0);
   \draw[ultra thick,-{Stealth[length=3mm]}] (-2.45,0) -- (-2.35,0);
   \draw[ultra thick,-{Stealth[length=3mm]}] (3.55,0) -- (3.65,0);
   \draw[ultra thick,-{Stealth[length=3mm]}] (-3.45,0) -- (-3.35,0);
   \draw[ultra thick,-{Stealth[length=3mm]}] (1.5,-0.5) -- (1.6,-0.4);
   \draw[ultra thick,-{Stealth[length=3mm]}]  (-1.5,-0.5) -- (-1.4,-0.6);
   \draw[ultra thick,-{Stealth[length=3mm]}]  (1.5,0.5) -- (1.6,0.4);
   \draw[ultra thick,-{Stealth[length=3mm]}]  (-1.5,0.5) -- (-1.4,0.6);
   \draw[ultra thick,-{Stealth[length=3mm]}]  (-1.55,-1.55) -- (-1.45,-1.45);
   \draw[ultra thick,-{Stealth[length=3mm]}]  (1.55,-1.55) -- (1.65,-1.65);
   \draw[ultra thick,-{Stealth[length=3mm]}]  (-1.55,1.55) -- (-1.45,1.45);
   \draw[ultra thick,-{Stealth[length=3mm]}]  (1.55,1.55) -- (1.65,1.65);

    \draw (-0.5,1.4) node {\huge $a$};
    \draw (-1.75,0.85) node {\huge $b$};
    \draw (-2.2,2.2) node {\huge $i$};
    \draw (-2.5,0.5) node {\huge $k$};
    \draw (-3.5,0.55) node {\huge $j$};
    \draw (-1.85,-0.75) node {\huge $c$};
    \draw (-2.2,-2.2) node {\huge $l$};
    \draw (-0.5,-1.4) node {\huge $d$};
    \draw (0.5,-1.5) node {\huge $e$};
    \draw (2.2,-2.3) node {\huge $m$};
    \draw (1.82,-0.75) node {\huge $f$};
    \draw (1.85,0.85) node {\huge $g$};
    \draw (2.5,0.4) node {\huge $n$};
    \draw (3.5,0.4) node {\huge $o$};
    \draw (2.3,2.1) node {\huge $p$};
    \draw (0.5,1.5) node {\huge $h$};
\end{tikzpicture}}}

\newcommand{\latticeplaquette}{\scalebox{0.45}{\begin{tikzpicture}[baseline=-0.25cm]
   \draw[thick,orange] (0,0) ellipse (1.5cm and 0.75cm);

    \draw[ultra thick] (-2,0) -- (-1,-1) -- (1,-1) -- (2,0) -- (1,1) -- (-1,1) -- cycle;
    \draw[ultra thick] (-2,0) -- (-4,0);
    \draw[ultra thick] (-1,-1) -- (-2,-2);
    \draw[ultra thick] (1,-1) -- (2,-2);
    \draw[ultra thick] (2,0) -- (4,0);
    \draw[ultra thick] (1,1) -- (2,2);
    \draw[ultra thick] (-1,1) -- (-2,2);

    \draw[thick] (-3,0) -- (-3,-1);
    \draw[thick] (0,1) -- (0,0.85);
    \draw[thick] (0,0.65) -- (0,0);
    \draw[thick] (0,-1) -- (0,-2);
    \draw[thick] (3,0) -- (3,-1);
    
    \draw[orange] (-0.75,0.25) node {\huge $\alpha$};
    \draw[ultra thick, orange,-{Stealth[length=3mm]}] (-0.85,0.62) -- (-0.94,0.58);
    
   \draw[ultra thick,-{Stealth[length=3mm]}] (-0.45,1) -- (-0.35,1);
   \draw[ultra thick,-{Stealth[length=3mm]}] (0.55,1) -- (0.65,1);
   \draw[ultra thick,-{Stealth[length=3mm]}] (-0.45,-1) -- (-0.35,-1);
   \draw[ultra thick,-{Stealth[length=3mm]}] (0.55,-1) -- (0.65,-1);
   \draw[ultra thick,-{Stealth[length=3mm]}] (2.55,0) -- (2.65,0);
   \draw[ultra thick,-{Stealth[length=3mm]}] (-2.45,0) -- (-2.35,0);
   \draw[ultra thick,-{Stealth[length=3mm]}] (3.55,0) -- (3.65,0);
   \draw[ultra thick,-{Stealth[length=3mm]}] (-3.45,0) -- (-3.35,0);
   \draw[ultra thick,-{Stealth[length=3mm]}] (1.5,-0.5) -- (1.6,-0.4);
   \draw[ultra thick,-{Stealth[length=3mm]}]  (-1.5,-0.5) -- (-1.4,-0.6);
   \draw[ultra thick,-{Stealth[length=3mm]}]  (1.5,0.5) -- (1.6,0.4);
   \draw[ultra thick,-{Stealth[length=3mm]}]  (-1.5,0.5) -- (-1.4,0.6);
   \draw[ultra thick,-{Stealth[length=3mm]}]  (-1.55,-1.55) -- (-1.45,-1.45);
   \draw[ultra thick,-{Stealth[length=3mm]}]  (1.55,-1.55) -- (1.65,-1.65);
   \draw[ultra thick,-{Stealth[length=3mm]}]  (-1.55,1.55) -- (-1.45,1.45);
   \draw[ultra thick,-{Stealth[length=3mm]}]  (1.55,1.55) -- (1.65,1.65);

    \draw (-0.5,1.4) node {\huge $a$};
    \draw (-1.75,0.85) node {\huge $b$};
    \draw (-2.2,2.2) node {\huge $i$};
    \draw (-2.5,0.5) node {\huge $k$};
    \draw (-3.5,0.55) node {\huge $j$};
    \draw (-1.85,-0.75) node {\huge $c$};
    \draw (-2.2,-2.2) node {\huge $l$};
    \draw (-0.5,-1.4) node {\huge $d$};
    \draw (0.5,-1.5) node {\huge $e$};
    \draw (2.2,-2.3) node {\huge $m$};
    \draw (1.82,-0.75) node {\huge $f$};
    \draw (1.85,0.85) node {\huge $g$};
    \draw (2.5,0.4) node {\huge $n$};
    \draw (3.5,0.4) node {\huge $o$};
    \draw (2.3,2.1) node {\huge $p$};
    \draw (0.5,1.5) node {\huge $h$};
\end{tikzpicture}}}

\newcommand{\latticeplaquettefused}{\scalebox{0.45}{\begin{tikzpicture}[baseline=-0.25cm]

    \draw[ultra thick,cyan] (-2,0) -- (-1,-1) -- (1,-1) -- (2,0) -- (1,1) -- (-1,1) -- cycle;
    \draw[ultra thick] (-2,0) -- (-4,0);
    \draw[ultra thick] (-1,-1) -- (-2,-2);
    \draw[ultra thick] (1,-1) -- (2,-2);
    \draw[ultra thick] (2,0) -- (4,0);
    \draw[ultra thick] (1,1) -- (2,2);
    \draw[ultra thick] (-1,1) -- (-2,2);

    \draw[thick] (-3,0) -- (-3,-1);
    \draw[thick] (0,1) -- (0,0);
    \draw[thick] (0,-1) -- (0,-2);
    \draw[thick] (3,0) -- (3,-1);
    
   \draw[ultra thick,cyan,-{Stealth[length=3mm]}] (-0.45,1) -- (-0.35,1);
   \draw[ultra thick,cyan,-{Stealth[length=3mm]}] (0.55,1) -- (0.65,1);
   \draw[ultra thick,cyan,-{Stealth[length=3mm]}] (-0.45,-1) -- (-0.35,-1);
   \draw[ultra thick,cyan,-{Stealth[length=3mm]}] (0.55,-1) -- (0.65,-1);
   \draw[ultra thick,-{Stealth[length=3mm]}] (2.55,0) -- (2.65,0);
   \draw[ultra thick,-{Stealth[length=3mm]}] (-2.45,0) -- (-2.35,0);
   \draw[ultra thick,-{Stealth[length=3mm]}] (3.55,0) -- (3.65,0);
   \draw[ultra thick,-{Stealth[length=3mm]}] (-3.45,0) -- (-3.35,0);
   \draw[ultra thick,cyan,-{Stealth[length=3mm]}] (1.5,-0.5) -- (1.6,-0.4);
   \draw[ultra thick,cyan,-{Stealth[length=3mm]}]  (-1.5,-0.5) -- (-1.4,-0.6);
   \draw[ultra thick,cyan,-{Stealth[length=3mm]}]  (1.5,0.5) -- (1.6,0.4);
   \draw[ultra thick,cyan,-{Stealth[length=3mm]}]  (-1.5,0.5) -- (-1.4,0.6);
   \draw[ultra thick,-{Stealth[length=3mm]}]  (-1.55,-1.55) -- (-1.45,-1.45);
   \draw[ultra thick,-{Stealth[length=3mm]}]  (1.55,-1.55) -- (1.65,-1.65);
   \draw[ultra thick,-{Stealth[length=3mm]}]  (-1.55,1.55) -- (-1.45,1.45);
   \draw[ultra thick,-{Stealth[length=3mm]}]  (1.55,1.55) -- (1.65,1.65);

    \draw (-0.5,1.5) node {\huge $a'$};
    \draw (-1.75,0.85) node {\huge $b'$};
    \draw (-2.2,2.2) node {\huge $i$};
    \draw (-2.5,0.5) node {\huge $k$};
    \draw (-3.5,0.55) node {\huge $j$};
    \draw (-1.9,-0.8) node {\huge $c'$};
    \draw (-2.2,-2.2) node {\huge $l$};
    \draw (-0.5,-1.5) node {\huge $d'$};
    \draw (0.5,-1.5) node {\huge $e'$};
    \draw (2.2,-2.3) node {\huge $m$};
    \draw (1.82,-0.75) node {\huge $f'$};
    \draw (1.85,0.95) node {\huge $g'$};
    \draw (2.5,0.4) node {\huge $n$};
    \draw (3.5,0.4) node {\huge $o$};
    \draw (2.3,2.1) node {\huge $p$};
    \draw (0.5,1.5) node {\huge $h'$};
\end{tikzpicture}}}

\newcommand{\xlattice}{\scalebox{0.45}{\begin{tikzpicture}[baseline=-0.75cm]

    \draw[ultra thick] (-2,0) -- (-1,-1) -- (1,-1);
    \draw[ultra thick] (-1,1) -- (-2,0);
    \draw[ultra thick] (-2,0) -- (-4,0);
    \draw[ultra thick] (-1,-1) -- (-2,-2);

    \draw[thick] (-3,0) -- (-3,-1);
    \draw[thick] (0,-1) -- (0,-2);
 
   \draw[ultra thick,-{Stealth[length=3mm]}] (-0.45,-1) -- (-0.35,-1);
   \draw[ultra thick,-{Stealth[length=3mm]}] (0.55,-1) -- (0.65,-1);
   \draw[ultra thick,-{Stealth[length=3mm]}] (-2.45,0) -- (-2.35,0);
   \draw[ultra thick,-{Stealth[length=3mm]}] (-3.45,0) -- (-3.35,0);
   \draw[ultra thick,-{Stealth[length=3mm]}]  (-1.5,-0.5) -- (-1.4,-0.6);
   \draw[ultra thick,-{Stealth[length=3mm]}]  (-1.5,0.5) -- (-1.4,0.6);
   \draw[ultra thick,-{Stealth[length=3mm]}]  (-1.55,-1.55) -- (-1.45,-1.45);

    \draw (-0.7,1.1) node {\huge $b$};
    \draw (-2.5,0.5) node {\huge $k$};
    \draw (-3.5,0.55) node {\huge $j$};
    \draw (-1.2,-0.2) node {\huge $c$};
    \draw (-2.2,-2.2) node {\huge $l$};
    \draw (-0.5,-0.45) node {\huge $d$};
    \draw (0.5,-0.55) node {\huge $e$};
\end{tikzpicture}}}

\newcommand{\ylattice}{\scalebox{0.45}{\begin{tikzpicture}[baseline=0.25cm]
    \draw[ultra thick] (1,1) -- (-1,1) -- (-2,0) -- (-1,-1);
    \draw[ultra thick] (-2,0) -- (-4,0);
    \draw[ultra thick] (-1,1) -- (-2,2);

    \draw[thick] (-3,0) -- (-3,-1);
    \draw[thick] (0,1) -- (0,0);
  
   \draw[ultra thick,-{Stealth[length=3mm]}] (-0.45,1) -- (-0.35,1);
   \draw[ultra thick,-{Stealth[length=3mm]}] (0.55,1) -- (0.65,1);
   \draw[ultra thick,-{Stealth[length=3mm]}] (-2.45,0) -- (-2.35,0);
   \draw[ultra thick,-{Stealth[length=3mm]}] (-3.45,0) -- (-3.35,0);
   \draw[ultra thick,-{Stealth[length=3mm]}]  (-1.5,-0.5) -- (-1.4,-0.6);
   \draw[ultra thick,-{Stealth[length=3mm]}]  (-1.5,0.5) -- (-1.4,0.6);
   \draw[ultra thick,-{Stealth[length=3mm]}]  (-1.55,1.55) -- (-1.45,1.45);

    \draw (-0.5,1.4) node {\huge $a$};
    \draw (-1.75,0.85) node {\huge $b$};
    \draw (-2.2,2.2) node {\huge $i$};
    \draw (-2.5,0.5) node {\huge $k$};
    \draw (-3.5,0.55) node {\huge $j$};
    \draw (-0.8,-1.1) node {\huge $c$};
    \draw (0.5,1.5) node {\huge $h$};
\end{tikzpicture}}}

\newcommand{\ylatticehopping}{\scalebox{0.45}{\begin{tikzpicture}[baseline=0.25cm]
   \draw[thick,orange] (-3,-0.5) -- (-1.98,-0.16);
   \draw[thick,orange] (0,0.5) -- (-1.78,-0.095);
   
   \draw[thick,orange,-{Stealth[length=3mm]}] (-1.4,0.025) -- (-1.3,0.057);

    \draw[ultra thick] (1,1) -- (-1,1) -- (-2,0) -- (-1,-1);
    \draw[ultra thick] (-2,0) -- (-4,0);
    \draw[ultra thick] (-1,1) -- (-2,2);

    \draw[thick] (-3,0) -- (-3,-1);
    \draw[thick] (0,1) -- (0,0);
    
    \node[orange] at (-0.85,-0.2) {\huge $\alpha$};
  
   \draw[ultra thick,-{Stealth[length=3mm]}] (-0.45,1) -- (-0.35,1);
   \draw[ultra thick,-{Stealth[length=3mm]}] (0.55,1) -- (0.65,1);
   \draw[ultra thick,-{Stealth[length=3mm]}] (-2.45,0) -- (-2.35,0);
   \draw[ultra thick,-{Stealth[length=3mm]}] (-3.45,0) -- (-3.35,0);
   \draw[ultra thick,-{Stealth[length=3mm]}]  (-1.5,-0.5) -- (-1.4,-0.6);
   \draw[ultra thick,-{Stealth[length=3mm]}]  (-1.5,0.5) -- (-1.4,0.6);
   \draw[ultra thick,-{Stealth[length=3mm]}]  (-1.55,1.55) -- (-1.45,1.45);

    \draw (-0.5,1.4) node {\huge $a$};
    \draw (-1.75,0.85) node {\huge $b$};
    \draw (-2.2,2.2) node {\huge $i$};
    \draw (-2.5,0.5) node {\huge $k$};
    \draw (-3.5,0.55) node {\huge $j$};
    \draw (-0.8,-1.1) node {\huge $c$};
    \draw (0.5,1.5) node {\huge $h$};
\end{tikzpicture}}}

\newcommand{\xlatticehopping}{\scalebox{0.45}{\begin{tikzpicture}[baseline=-0.75cm]
   \draw[thick,orange] (-3,-0.6) -- (-1.3,-1.165);
   \draw[thick,orange] (0,-1.6) -- (-1.1,-1.235);
   
    \draw[thick,orange,-{Stealth[length=3mm]}] (-1.8,-1) -- (-1.7,-1.04);
   
    \draw[ultra thick] (-2,0) -- (-1,-1) -- (1,-1);
    \draw[ultra thick] (-1,1) -- (-2,0);
    \draw[ultra thick] (-2,0) -- (-4,0);
    \draw[ultra thick] (-1,-1) -- (-2,-2);

    \draw[thick] (-3,0) -- (-3,-1);
    \draw[thick] (0,-1) -- (0,-2);

    \node[orange] at (-2.3,-1.3) {\huge $\alpha$};
 
   \draw[ultra thick,-{Stealth[length=3mm]}] (-0.45,-1) -- (-0.35,-1);
   \draw[ultra thick,-{Stealth[length=3mm]}] (0.55,-1) -- (0.65,-1);
   \draw[ultra thick,-{Stealth[length=3mm]}] (-2.45,0) -- (-2.35,0);
   \draw[ultra thick,-{Stealth[length=3mm]}] (-3.45,0) -- (-3.35,0);
   \draw[ultra thick,-{Stealth[length=3mm]}]  (-1.5,-0.5) -- (-1.4,-0.6);
   \draw[ultra thick,-{Stealth[length=3mm]}]  (-1.5,0.5) -- (-1.4,0.6);
   \draw[ultra thick,-{Stealth[length=3mm]}]  (-1.55,-1.55) -- (-1.45,-1.45);

    \draw (-0.7,1.1) node {\huge $b$};
    \draw (-2.5,0.5) node {\huge $k$};
    \draw (-3.5,0.55) node {\huge $j$};
    \draw (-1.2,-0.2) node {\huge $c$};
    \draw (-2.2,-2.2) node {\huge $l$};
    \draw (-0.5,-0.45) node {\huge $d$};
    \draw (0.5,-0.55) node {\huge $e$};
\end{tikzpicture}}}

\newcommand{\ylatticehoppingfused}{\scalebox{0.45}{\begin{tikzpicture}[baseline=0.25cm]
    \draw[ultra thick] (1,1) -- (0,1);
    \draw[ultra thick,cyan] (0,1) -- (-1,1);
    \draw[ultra thick,cyan] (-1,1) -- (-2,0);
    \draw[ultra thick] (-2,0) -- (-1,-1);
    \draw[ultra thick,cyan] (-2,0) -- (-3,0);
    \draw[ultra thick] (-3,0) -- (-4,0);
    \draw[ultra thick] (-1,1) -- (-2,2);

    \draw[thick] (-3,0) -- (-3,-1);
    \draw[thick] (0,1) -- (0,0);
  
   \draw[ultra thick,cyan,-{Stealth[length=3mm]}] (-0.45,1) -- (-0.35,1);
   \draw[ultra thick,-{Stealth[length=3mm]}] (0.55,1) -- (0.65,1);
   \draw[ultra thick,cyan,-{Stealth[length=3mm]}] (-2.45,0) -- (-2.35,0);
   \draw[ultra thick,-{Stealth[length=3mm]}] (-3.45,0) -- (-3.35,0);
   \draw[ultra thick,-{Stealth[length=3mm]}]  (-1.5,-0.5) -- (-1.4,-0.6);
   \draw[ultra thick,cyan,-{Stealth[length=3mm]}]  (-1.5,0.5) -- (-1.4,0.6);
   \draw[ultra thick,-{Stealth[length=3mm]}]  (-1.55,1.55) -- (-1.45,1.45);

    \draw (-0.5,1.5) node {\huge $a'$};
    \draw (-1.75,0.85) node {\huge $b'$};
    \draw (-2.2,2.2) node {\huge $i$};
    \draw (-2.5,0.55) node {\huge $k'$};
    \draw (-3.5,0.55) node {\huge $j$};
    \draw (-0.8,-1.1) node {\huge $c$};
    \draw (0.5,1.5) node {\huge $h$};
\end{tikzpicture}}}

\newcommand{\xlatticehoppingfused}{\scalebox{0.45}{\begin{tikzpicture}[baseline=-0.75cm]
    \draw[ultra thick,cyan] (-2,0) -- (-1,-1) -- (0,-1);
    \draw[ultra thick] (0,-1) -- (1,-1);
    \draw[ultra thick] (-1,1) -- (-2,0);
    \draw[ultra thick,cyan] (-2,0) -- (-3,0);
    \draw[ultra thick] (-3,0) -- (-4,0);
    \draw[ultra thick] (-1,-1) -- (-2,-2);

    \draw[thick] (-3,0) -- (-3,-1);
    \draw[thick] (0,-1) -- (0,-2);
 
   \draw[ultra thick,cyan,-{Stealth[length=3mm]}] (-0.45,-1) -- (-0.35,-1);
   \draw[ultra thick,-{Stealth[length=3mm]}] (0.55,-1) -- (0.65,-1);
   \draw[ultra thick,cyan,-{Stealth[length=3mm]}] (-2.45,0) -- (-2.35,0);
   \draw[ultra thick,-{Stealth[length=3mm]}] (-3.45,0) -- (-3.35,0);
   \draw[ultra thick,cyan,-{Stealth[length=3mm]}]  (-1.5,-0.5) -- (-1.4,-0.6);
   \draw[ultra thick,-{Stealth[length=3mm]}]  (-1.5,0.5) -- (-1.4,0.6);
   \draw[ultra thick,-{Stealth[length=3mm]}]  (-1.55,-1.55) -- (-1.45,-1.45);

    \draw (-0.7,1.1) node {\huge $b$};
    \draw (-2.5,0.5) node {\huge $k'$};
    \draw (-3.5,0.55) node {\huge $j$};
    \draw (-1.15,-0.15) node {\huge $c'$};
    \draw (-2.2,-2.2) node {\huge $l$};
    \draw (-0.5,-0.45) node {\huge $d'$};
    \draw (0.5,-0.55) node {\huge $e$};
\end{tikzpicture}}}

\newcommand{\latticemass}{\scalebox{0.75}{\begin{tikzpicture}[baseline=-0.1cm]
 \draw[thick] (0,-0.5) -- (0,0.5);
 \draw[ thick] (-1,0.5) -- (1,0.5);
 
  \draw[ thick,-{Stealth[length=2mm]}] (-0.55,0.5) -- (-0.45,0.5);
  \draw[ thick,-{Stealth[length=2mm]}] (0.5,0.5) -- (0.6,0.5);

    \draw (-0.5,0.15) node {\Large $a$};
    \draw (0.5,0.2) node {\Large $b$};
\end{tikzpicture}}}

\newcommand{\latticemasstwist}{\scalebox{0.75}{\begin{tikzpicture}[baseline=-0.1cm]
 \draw[thick] (0,-0.5) -- (0,0.5);
 \draw[thick, orange] (-0.15,0.2) +(0:0.1) arc (110:270:0.1);
 \draw[thick, orange] (-0.12,0.2) +(0:0.2) arc (70:-90:0.2);

 \draw[thick] (-1,0.5) -- (1,0.5);
 
  \draw[ultra thick,-{Stealth[length=2mm]}] (-0.55,0.5) -- (-0.45,0.5);
  \draw[ultra thick,-{Stealth[length=2mm]}] (0.5,0.5) -- (0.6,0.5);

      \draw (-0.5,0.15) node {\Large $a$};
    \draw (0.5,0.2) node {\Large $b$};
     \node[orange] at (0.4,-0.2) {\large $\psi$};
\end{tikzpicture}}}

\newcommand{\latticeelectric}{\scalebox{1}{\begin{tikzpicture}[baseline=-0.08cm]
\draw[thick] (-0.5,-0.5) -- (0.5,0.5);
\draw (0.18,0.4) node {\small $b$};

 \draw[thick,-{Stealth[length=1.5mm]}] (-0.2,-0.2) -- (-0.1,-0.1);
\end{tikzpicture}}}

\newcommand{\latticeelectricloop}{\scalebox{1}{\begin{tikzpicture}[baseline=-0.08cm]
\draw[thick] (-0.5,-0.5) -- (0.5,0.5);
\draw[orange] (0,0) +(70:0.1) arc (70:380:0.1);
\draw[orange] (0.2,-0.1) node {\small $c$};
\draw (0.18,0.4) node {\small $b$};

 \draw[thick,-{Stealth[length=1.5mm]}] (-0.2,-0.2) -- (-0.1,-0.1);
\draw[orange,-{Stealth[length=0.8mm]}] (-0.095,0.02) -- (-0.1,-0.02);
\end{tikzpicture}}}

\newcommand{\resolveloopinitial}{\scalebox{1}{\begin{tikzpicture}[baseline=0]
    \draw[thick,-{Stealth[length=2mm]}] (0,-0.75) -- (0,-0.65);
    \draw[thick] (0,-1) -- (0,-0.5);

    \draw[thick] (0,-0.5) to[out=130, in=230, looseness=1] (0,0.5);
    \draw[thick] (0,-0.5) to[out=50, in=310, looseness=1] (0,0.5);
    \draw[thick,-{Stealth[length=2mm]}] (-0.19,0) -- (-0.19,0.1);
    \draw[thick,-{Stealth[length=2mm]}] (0.19,0) -- (0.19,0.1);

    \draw[thick,-{Stealth[length=2mm]}] (0,0.75) -- (0,0.85);
    \draw[thick] (0,0.5) -- (0,1);
    
    \draw (-0.25,-0.75) node {$c$};
    \draw (-0.25,0.8) node {$c'$};
    \draw (-0.45,-0.05) node {$a$};
    \draw (0.4,0.0) node {$b$};
\end{tikzpicture}}}

\newcommand{\resolveloopfinal}{\scalebox{1}{\begin{tikzpicture}[baseline=0]
    \draw[thick,-{Stealth[length=2.2mm]}] (0,0) -- (0,0.1);
    \draw[thick] (0,-0.95) -- (0,0.95);

    \draw (0.25,-0.05) node {$c$};
\end{tikzpicture}}}

\newcommand{\resolveidentityinitial}{\scalebox{1}{\begin{tikzpicture}[baseline=0]
     \draw[thick,-{Stealth[length=2.2mm]}] (-0.19,0) -- (-0.19,0.1);
    \draw[thick] (-0.19,-1) -- (-0.19,1);
     \draw[thick,-{Stealth[length=2.2mm]}] (0.19,0) -- (0.19,0.1);
    \draw[thick] (0.19,-1) -- (0.19,1);

    \draw (-0.45,-0.05) node {$a$};
    \draw (0.4,0) node {$b$};
\end{tikzpicture}}}

\newcommand{\resolveidentityfinal}{\scalebox{1}{\begin{tikzpicture}[baseline=0]
     \draw[thick,-{Stealth[length=2mm]}] (0,0) -- (0,0.1);
    \draw[thick] (0,-0.5) -- (0,0.5);

     \draw[thick,-{Stealth[length=2mm]}] (0.433,0.75) -- (0.533,0.81);
    \draw[thick] (0.866,1) -- (0,0.5);

    \draw[thick,-{Stealth[length=2mm]}] (-0.433,0.75) -- (-0.533,0.81);
    \draw[thick] (-0.866,1) -- (0,0.5);

    \draw[thick,-{Stealth[length=2mm]}] (0.533,-0.81) -- (0.433,-0.75);
    \draw[thick] (0.866,-01) -- (0,-0.5);

    \draw[thick,-{Stealth[length=2mm]}] (-0.533,-0.81) -- (-0.433,-0.75);
    \draw[thick] (-0.866,-1) -- (0,-0.5);

    \draw (0.2,0) node {$c$};
    \draw (-0.5,0.5) node {$a$};
    \draw (-0.5,-0.55) node {$a$};
    \draw (0.5,0.55) node {$b$};
    \draw (0.5,-0.5) node {$b$};
\end{tikzpicture}}}

\newcommand{\Fmoveinitial}{\scalebox{1}{\begin{tikzpicture}[baseline=0]
     \draw[thick,-{Stealth[length=2mm]}] (0,-0.75) -- (0,-0.65);
    \draw[thick] (0,-1) -- (0,-0.5);

     \draw[thick,-{Stealth[length=2mm]}] (0.48,0.225) -- (0.57,0.35);
    \draw[thick] (1,1) -- (0,-0.5);

    \draw[thick,-{Stealth[length=2mm]}] (-0.83,0.75) -- (-0.87,0.81);
    \draw[thick] (-1,1) -- (0,-0.5);

    \draw[thick,-{Stealth[length=2mm]}] (-0.17,0.75) -- (-0.13,0.81);
    \draw[thick] (-0.5,0.25) -- (0,1);

    \draw[thick,-{Stealth[length=2mm]}] (-0.33,0) -- (-0.37,0.06);

    \draw (-0.55,-0.15) node {$e$};
    \draw (-1.1,1.1) node {$a$};
    \draw (0.05,1.15) node {$b$};
    \draw (1.1,1.1) node {$c$};
    \draw (0,-1.2) node {$d$};
\end{tikzpicture}}}

\newcommand{\Fmovefinal}{\scalebox{1}{\begin{tikzpicture}[baseline=0]
     \draw[thick,-{Stealth[length=2mm]}] (0,-0.75) -- (0,-0.65);
    \draw[thick] (0,-1) -- (0,-0.5);

     \draw[thick,-{Stealth[length=2mm]}] (-0.48,0.225) -- (-0.57,0.35);
    \draw[thick] (1,1) -- (0,-0.5);

    \draw[thick,-{Stealth[length=2mm]}] (0.83,0.75) -- (0.87,0.81);
    \draw[thick] (-1,1) -- (0,-0.5);

    \draw[thick,-{Stealth[length=2mm]}] (0.17,0.75) -- (0.13,0.81);
    \draw[thick] (0.5,0.25) -- (0,1);

    \draw[thick,-{Stealth[length=2mm]}] (0.33,0) -- (0.37,0.06);

    \draw (0.5,-0.15) node {$f$};
    \draw (-1.1,1.1) node {$a$};
    \draw (-0.05,1.15) node {$b$};
    \draw (1.1,1.1) node {$c$};
    \draw (0,-1.2) node {$d$};
\end{tikzpicture}}}

\newcommand{\Rmoveinitial}{\scalebox{1}{\begin{tikzpicture}[baseline=0]
     \draw[thick,-{Stealth[length=2mm]}] (0,-0.35) -- (0,-0.25);
    \draw[thick] (0,-0.75) -- (0,0);

     \draw[thick] (0,0) to[out=170, in=220, looseness=1.2] (0.5,1);
     \draw[thick] (0,0) to[out=10, in=310, looseness=1.2] (0.05,0.45);

     \draw[thick] (-0.1,0.6) -- (-0.5,1);

     \draw[thick,-{Stealth[length=2mm]}] (-0.35,0.85) -- (-0.41,0.92);
     \draw[thick,-{Stealth[length=2mm]}] (0.3,0.82) -- (0.36,0.88);

    \draw (-0.6,1.2) node {$b$};
    \draw (0.6,1.15) node {$c$};
    \draw (-0.25,-0.4) node {$a$};
\end{tikzpicture}}}

\newcommand{\Rmovefinal}{\scalebox{1}{\begin{tikzpicture}[baseline=0]
     \draw[thick,-{Stealth[length=2mm]}] (0,-0.35) -- (0,-0.25);
    \draw[thick] (0,-0.75) -- (0,0);

     \draw[thick] (0,0) -- (0.5,1);
     \draw[thick] (0,0) -- (-0.5,1);

     \draw[thick,-{Stealth[length=2mm]}] (-0.25,0.5) -- (-0.3,0.6);
     \draw[thick,-{Stealth[length=2mm]}] (0.25,0.5) -- (0.3,0.6);

    \draw (-0.6,1.2) node {$b$};
    \draw (0.6,1.15) node {$c$};
    \draw (-0.25,-0.4) node {$a$};
\end{tikzpicture}}}

\begin{document}

\title{Quantum simulation of lattice gauge theories coupled to fermionic matter via anyonic regularization}

%Affiliations
%%%%%%%%%%%%%%%%%%%%%%%%%%%%%%%%%%%%%%%%%%%%%%%%%%%%%%%%%%%%%%%%%%%%
%%%%%%%%%%%%%%%%%%%%%%%%%%%%%%%%%%%%%%%%%%%%%%%%%%%%%%%%%%%%%%%%%%%%
\author{Mason L. \surname{Rhodes}}
\email[]{mlrhod@sandia.gov}
\affiliation{Center for Computing Research,
             Sandia National Laboratories,
             Albuquerque, NM, 87185, USA}
\affiliation{Center for Quantum Information and Control,
             University of New Mexico,
             Albuquerque, NM, 87131, USA}
\affiliation{Department of Physics and Astronomy,
             University of New Mexico,
             Albuquerque, NM, 87131, USA}

\author{Shivesh Pathak}
\affiliation{Quantum Algorithms and Applications Collaboratory, Sandia National Laboratories, Albuquerque, NM, USA}

\author{Riley W. Chien}
\affiliation{Center for Computing Research,
             Sandia National Laboratories,
             Albuquerque, NM, 87185, USA}
\affiliation{Quantum Algorithms and Applications Collaboratory, Sandia National Laboratories, Albuquerque, NM, USA}
%%%%%%%%%%%%%%%%%%%%%%%%%%%%%%%%%%%%%%%%%%%%%%%%%%%%%%%%%%%%%%%%%%%%
%%%%%%%%%%%%%%%%%%%%%%%%%%%%%%%%%%%%%%%%%%%%%%%%%%%%%%%%%%%%%%%%%%%%

\date{\today}

\begin{abstract}
The optimal regularization of infinite-dimensional gauge-field degrees of freedom is a central open problem in the simulation of lattice gauge theories on quantum computers. Here, we consider regularizing the gauge field by replacing the gauge group $G$ with a braided fusion category whose objects correspond to Wilson lines of the associated Chern-Simons theory $G_k$, with the level $k$ serving as the regularization parameter. We demonstrate how to couple these regularized gauge groups to fermionic matter using the framework of fusion surface models, which treats matter and gauge field excitations as interacting anyons. We then address the simulation of the regularized Hamiltonian, in the Kogut-Susskind formulation, on fault-tolerant quantum computers.
We provide explicit quantum circuit constructions for implementing the primitive gates in this model, the $F$ and $R$ symbols, for $U(1)_k$ and $SU(2)_k$ anyon theories.
\end{abstract}

\maketitle
\bfseries
\tableofcontents
\normalfont
\hrulefill

%%%%%%%%%%%%%%%%%%%%%%%%%%%%%%%%%%%%%%%%%%%%%%%%%%%%%%%%%%%%%%%%%%%%
%%%%%%%%%%%%%%%%%%%%%%%%%%%%%%%%%%%%%%%%%%%%%%%%%%%%%%%%%%%%%%%%%%%%

\section{Introduction}
\label{sec:Introduction}

%%%%%%%%%%%%%%%%%%%%%%%%%%%%%%%%%%%%%%%%%%%%%%%%%%%%%%%%%%%%%%%%%%%%
%%%%%%%%%%%%%%%%%%%%%%%%%%%%%%%%%%%%%%%%%%%%%%%%%%%%%%%%%%%%%%%%%%%%

The simulation of lattice gauge theories (LGTs) on quantum computers has received significant attention as a potential application of practical quantum advantage. Following the foundational work of Byrnes and Yamamoto~\cite{Byrnes:2006a}, particularly in the Hamiltonian formulation of Kogut and Susskind~\cite{Kogut:1975a}, 
several advances have been made in both the time- and space-efficiency of Hamiltonian simulation algorithms for LGTs, including improvements to the representation~\cite{Lamm:2019a, Alexandru:2019a, Ji:2020a, Raychowdhury:2020a, Raychowdhury:2020b, Buser:2021a, Wiese:2021a, Carena:2022a, Ji:2023a, Zache:2023b, Hayata:2023a, Bergner:2024a, Assi:2024a,modi2026largenctruncationssunc} and encoding~\cite{Zohar:2015a, Zohar:2017a, Klco:2019a, Zohar:2019a, Lamm:2020a, Bender:2020a, Liu:2020a, Barata:2021a, Ciavarella:2021a, Davoudi:2021a} of the Hamiltonian itself, as well as to explicit algorithmic implementations~\cite{Banerjee:2012a, Banerjee:2013a, Zohar:2013a, Zohar:2013b, Zohar:2013c, Tagliacozzo:2013a, Mezzacapo:2015a, Zohar:2017b, Zache:2018a, Bender:2018a, Klco:2020a, Shaw:2020a, Carena:2021a, Haase:2021a, Paulson:2021a, Alam:2022a, Alexandru:2022a, Kan:2022a, Gustafson:2022a, Dasgupta:2022a, Gustafson:2024a, Gustafson:2024b, Lamm:2024a, Rhodes:2024a, Halimeh:2024a, Hanada:2025a, Hayata:2025a}. For a comprehensive review of the simulation of quantum field theories, see Ref.~\cite{Bauer:2023a}.

A significant open problem in the simulation of LGTs is the regularization of the Hilbert space of the gauge fields. 
The gauge groups relevant to particle physics are Lie groups, and thereby require a regularization scheme to map onto the finite dimensional Hilbert spaces accessible on quantum computers. 

The most common regularization scheme is direct truncation~\cite{Byrnes:2006a, Zohar:2015a, Zohar:2017a, Klco:2019a, Zohar:2019a, Klco:2020a, Kan:2022a, Rhodes:2024a}. Indexing the countably infinite irreducible representations (irreps) of the gauge group by integers $\lambda\in\mathbb{Z}$, one fixes some cutoff $\Lambda$ and retains only representations indexed by $\lambda\in[-\Lambda,\Lambda]$. Choosing $\Lambda$ significantly larger than the energy scale of interest permits an accurate simulation of the low-energy physics of the system. 
However, since irreps for $\lambda > \Lambda$ are undefined, it is unclear how to systematically bound the truncation error in the continuum limit, the latter being required to recover the continuous spacetime in which observables become physical.
Additionally, it is not known, for a given application, how large $\Lambda$ must be to obtain physically meaningful results without relying on experimentally verified properties of the system, reducing the predictive power of this model.
We also note that direct truncation of the gauge group $SU(N)$ leads to untenable quantum circuit realizations for general $N\geq 2$~\cite{Rhodes:2024a}.

An alternative regularization replaces the full gauge group with one of its discrete subgroups~\cite{Lamm:2019a, Alexandru:2019a, Ji:2020a, Ji:2023a, Assi:2024a}. Since the discrete subgroups of a Lie group preserve its symmetries, this approach allows one to systematically bound the error in the continuum limit, although it is only valid in the limited regimes below the \emph{freezing couplings} of the discrete subgroups, which in practice, can be quite small~\cite{Gustafson:2022a, Gustafson:2024a, Gustafson:2024b}. Further, since every compact Lie group has a largest finite subgroup that approximates it, errors in simulation beyond these subgroups will be uncontrolled~\cite{Zache:2023b}.

More recently, a regularization scheme was proposed which aims to address both the continuum and group truncation limits. This approach replaces the full gauge group with a \emph{quantum group}, or equivalently, a $q$-deformation of the (universal enveloping algebra of the) associated Lie algebra~\cite{Zache:2023b,Hayata:2023a,Hayata:2025a}. Taking as input the quantum group $\mathcal{U}_q(G)$ of a gauge group $G$, the $q$-deformation has only a finite number of irreps when $q$ is a root of unity, and taking the limit $q\to 1$ recovers the full gauge group. Additionally, much like with discrete subgroups, one can systematically bound the error in the continuum limit as a function of the lattice spacing since the full gauge symmetries are retained in $\mathcal{U}_q(G)$. For these reasons, quantum group regularization is a promising candidate for efficient and accurate quantum simulation of LGTs.

Prior work in quantum group regularization, however, has three significant shortcomings.
Firstly, the coupling of fermionic matter to the regularized gauge field, which is essential in studying fundamental interactions relevant to particle physics and condensed matter.
To date, work has focused primarily on the simulation of pure gauge models without matter~\cite{Zache:2023b,Hayata:2023a,Hayata:2025a}, with a singular case where hard-core bosonic matter coupled to the regularized gauge field is considered~\cite{Zache:2023b}.
Secondly, that the typical Drinfeld-Jimbo construction for quantum groups is limited to semi-simple or simple gauge groups~\cite{Voigt2020}, a condition which is not satisfied by all gauge groups of interest, such as $U(N)$.
Thirdly, that the technique is restricted to lattices with two or fewer spatial dimensions.

In this work, we generalize quantum group constructions, developing a regularization scheme we call \emph{anyonic regularization}. Rather than using quantum groups, we replace the full gauge group $G$ with a braided fusion category (anyon model), whose objects correspond to Wilson lines of a Chern-Simons theory $G_k$, where the level $k$ serves as the tuning parameter, analogous to the deformation $q$, determining the number of irreps we keep. 
This immediately addresses the second shortcoming above, as level-$k$ Chern-Simons theories are well defined for non-semi-simple gauge groups, such as $U(N)_k$; we also note that the braided fusion category is equivalent to the quantum group construction in the case of (semi-)simple gauge groups like $SU(N)_k$. 
Using our novel regularization technique, we also provide, for the first time, an explicit construction of the anyonic-regularized Kogut-Susskind Hamiltonian coupled to fermionic matter, resolving the remaining shortcoming of quantum group constructions.
However, since we make use of anyonic objects, this technique is similarly restricted to two or fewer spatial dimensions.

To formalize our construction, we make use of a novel concept of \emph{fusion surface models}~\cite{Inamura:2024a}.
These models are defined on oriented trivalent lattices which take as input a braided fusion category\footnote{In full generality, the fusion surface model takes a fusion 2-category as input. 
Further, in the case we consider of the input being a braided fusion category, it would be more correct to say that we take the modules over the braided fusion category as the input. 
This lets one consider the membrane-like condensation defects. Since we do not make use of these, we will not explicitly use this language.}, and form a Hilbert space spanned by fusion diagrams, corresponding to labelings of edges by simple objects of the fusion category such that the associated fusion rules are obeyed at all vertices. 
The operators acting on this space take the form of Wilson loops and lines.
Resolution of the Wilson lines and loops into the lattice reveals an even more primitive structure, in terms of diagrammatic operators called $F$ and $R$ symbols.
We show that the regularized Hamiltonian expressed with these symbols takes the form of a linear combination of unitaries (LCU).
Thereby, any state-of-the-art Hamiltonian simulation algorithm, such as Trotterization~\cite{Childs:2021a} or qubitization~\cite{Tong:2022a}, can be utilized by querying circuit primitives which implement the diagrammatic $F$ and $R$ operators.

This article is structured as follows: In Sec.~\ref{sec:Preliminaries} we present the necessary background, including an introduction to fusion categories and anyon models in Sec.~\ref{sec:Fusion_Categories}, the canonical Kogut-Susskind Hamiltonian describing LGTs in Sec.~\ref{sec:KS_Ham}, and prior work on $q$-deformed LGTs without matter in Sec.~\ref{sec:Deformed-YM-Theory}. In Sec.~\ref{sec:Deformed_KS_Hamiltonian}, we develop the full anyonic-regularized Kogut-Susskind Hamiltonian, first by realizing purely-fermionic Hamiltonians in Sec.~\ref{sec:fermionic_Hamiltonians}, and then constructing operators which couple the fermionic matter to the anyonic-regularized gauge group in Sec.~\ref{sec:Constructing_Hamiltonian}. With the Hamiltonian in hand, we consider quantum simulation of $U(1)_k$ and $SU(2)_k$ LGTs by developing explicit quantum circuits for the primitive gates in this formalism, namely the $F$ and $R$ symbols of the anyon model, in Sec.~\ref{sec:Quantum_simulation}. Lastly, we summarize and provide future directions in Sec.~\ref{sec:Conclusion}.

%%%%%%%%%%%%%%%%%%%%%%%%%%%%%%%%%%%%%%%%%%%%%%%%%%%%%%%%%%%%%%%%%%%%
%%%%%%%%%%%%%%%%%%%%%%%%%%%%%%%%%%%%%%%%%%%%%%%%%%%%%%%%%%%%%%%%%%%%
\section{Preliminaries}
\label{sec:Preliminaries}

\subsection{Fusion Categories and Anyon Models}
\label{sec:Fusion_Categories}

%%%%%%%%%%%%%%%%%%%%%%%%%%%%%%%%%%%%%%%%%%%%%%%%%%%%%%%%%%%%%%%%%%%%

Here we provide a brief introduction to fusion categories and anyon models. In this work, we consider only braided fusion 1-categories, so we restrict our attention to these relevant properties. For additional details see Ref.~\cite[Sec. II]{Barkeshli:2019a} or Refs.~\cite{Wang:2010a,Simon:2023a}.

A braided fusion category $\mathcal{C}$ has objects labeled $a,b,c,\ldots\in\Obj(\mathcal{C})$ corresponding to topological excitations, or anyons. The category $\mathcal{C}$ is equipped with a fusion algebra which must be associative up to isomorphism and commutative\footnote{The additional constraint of commutativity is to accommodate braiding. A general unitary fusion category need only be associative up to isomorphism.}. The fusion algebra is defined by finite non-negative integer fusion coefficients $N_{ab}^c=N_{ba}^c$ satisfying
\al{\label{eq:fusion}
    a\times b=\sum_{c\in\mathcal{C}}N_{ab}^c c\,.
}
Throughout this work we consider only multiplicity-free fusion categories which further constrains the fusion coefficients such that $N_{ab}^c\in\{0,1\}$. This rules out the important gauge group $SU(3)$, however we do not foresee any major obstacle in this extension. Moreover, when $\mathcal{C}$ is Abelian, there is a unique fusion product $a\times b=c$.

The fusion algebra can be represented in a simple diagrammatic language. An anyon is represented by a worldline, or edge, carrying a label $a\in\mathcal{C}$, and the fusion rule of Eq.~\eqref{eq:fusion} is represented by trivalent vertices with incoming edges corresponding to the fusion or splitting of anyons. These trivalent vertices can be composed with each other by matching edges with equivalent anyons to construct arbitrary diagrams. Given such a diagram, there are a series of four operations one can perform on the lattice edges that leave the system invariant. The first move is the ability to remove closed loops in the diagrams, which corresponds to taking an inner product if the trivalent vertices are regarded as elements of the vector spaces $\Hom(a\times b,c)$:
\al{
    \resolveloopinitial = \delta_{c,c'}\sqrt{\frac{\text{d}_a\text{d}_b}{\text{d}_c}}\:\:\resolveloopfinal\,.
}
Here the coefficient $\text{d}_a$ is the \emph{quantum dimension} of particle $a$, a measure of the ``internal degree of freedom" carried by the anyons that ensures the planar diagrams are isotopy invariant. Using these, we can also define the total quantum dimension
\al{
    \mathcal{D}=\sqrt{\sum_{a\in\mathcal{C}} \text{d}_a^2}\,.
}

The second move allows for resolutions of the identity operator, which fuses and splits two anyons according to the fusion rules of Eq.~\eqref{eq:fusion}:
\al{
    \resolveidentityinitial = \sum_{c\in\mathcal{C}}\sqrt{\frac{\text{d}_c}{\text{d}_a\text{d}_b}}\:\:\resolveidentityfinal\,.
}
The third move corresponds to a unitary change of basis in the vector space $\Hom(a \times b \times c,d)$, which changes the order in which anyons are fused together. This move defines the so-called $F$ symbols of the anyon model:
\al{
    \Fmoveinitial = \sum_f [F_d^{abc}]_{ef}\:\:\Fmovefinal\,.
}
If it is possible to map one fusion diagram to another via two different sequences of these $F$ symbols, then those two sequences must be equivalent, a consequence of a nonlinear constraint known as the \emph{pentagon equation}~\cite{MacLane:1998a}.

While these first three moves are valid for any unitary fusion category, a fourth move arises for braided fusion categories, which exchanges the locations of two anyons. This move defines the so-called $R$ symbols of the anyon model:
\al{
    \Rmoveinitial = R_{a}^{bc}\:\:\Rmovefinal\,.
}
As before, since two sequences of $F$ and $R$ symbols that map an initial fusion diagram to the same final diagram should be equivalent, a new constraint equation is derived known as the \emph{hexagon equation}~\cite{MacLane:1998a}.
This equation ensures compatibility between fusion and braiding in the theory, and similarly, different consistent sets of $R$ symbols define different braided fusion categories, so the $R$ symbols must also be specified as part of the input data. Note that the elements of the $F$ and $R$ symbols are subject to gauge transformations and are not unique for a theory.

From these symbols one can then define various other properties of interest within the fusion category, including topological spin and the modular $S$ and $T$ matrices. The topological spin is
\al{
    \theta_a=\sum_{b\in\mathcal{C}} \frac{\text{d}_a}{\text{d}_b}R_b^{aa}\,,
}
which describes the phase accumulated from an anyon braiding with itself. Using these quantities, one can then define the modular matrices
\al{
    S_{ab}=\frac{1}{\mathcal{D}}\sum_{c\in\mathcal{C}}N_{\overline{a}b}^c\frac{\theta_c}{\theta_a\theta_b}\text{d}_c\,,\quad T_{ab}=\theta_{a}\delta_{ab}\,,
}
which, unlike the $F$ and $R$ symbols, are invariants of a braided fusion category.

%%%%%%%%%%%%%%%%%%%%%%%%%%%%%%%%%%%%%%%%%%%%%%%%%%%%%%%%%%%%%%%%%%%%
%%%%%%%%%%%%%%%%%%%%%%%%%%%%%%%%%%%%%%%%%%%%%%%%%%%%%%%%%%%%%%%%%%%%

\subsection{The Kogut-Susskind (KS) Hamiltonian}
\label{sec:KS_Ham}

In a seminal 1975 paper, Kogut and Susskind developed the now-standard Hamiltonian formulation of lattice gauge theories (LGTs)~\cite{Kogut:1975a}. The Hamiltonian is defined on a $d$-dimensional cubic spatial lattice consisting of $L^d$ vertices labeled by $\bm{v}\in\mathbb{Z}^d$. Fermionic matter is hosted on the vertices and the gauge-field degrees of freedom are hosted on edges. 

The Kogut-Susskind (KS) Hamiltonian can be explicitly written as
\al{
    H &= g_M H_M + g_K H_K + H_{YM}\,,
    \label{eq:KS_Hamiltonian}
}
where the coupling coefficients $g_M=m$ and $g_K=1/2a$ are defined in terms of the fermion mass $m$ and the lattice spacing $a$. The mass term $H_M$ and kinetic term $H_K$ act on the fermionic fields in the LGT and the Yang-Mills term $H_{YM}$ acts purely on the gauge fields, with each defined by
\al{
    H_M&=\sum_{\bm{v}} (-1)^{\bm{v}} a_{\bm{v}}^\dagger a_{\bm{v}}^\pd\,, \\
    H_K&=\sum_{\bm{e}} a_{\partial \bm{e}_1}^{\dag}W_{1}(\bm{e})a_{\partial\bm{e}_0}^\pd + \text{H.c.}\,, \\
    H_{YM} &= \frac{g^2}{2}\sum_{\bm{e}} E_{\bm{e}}^2 - \frac{1}{2a^2g^2}\sum_{\bm{p}} \Tr W_{1}(\partial \bm{p}) + W_{1}(\partial \bm{p})^{\dag}\,.
}
where $a_{\bm{v}}$ are the fermionic annihilation operators acting on each vertex $\bm{v}$.
In the mass term, $H_M$, the $(-1)^{\bm{v}}$ coefficient staggers fermionic matter and antimatter on the lattice, a result of splitting the Dirac spinors into their upper and lower components, which is necessary to mitigate fermion doubling. The operator $W_1(\bm{e})$ is the elementary Wilson link in the fundamental representation and the Wilson loop $\Tr W_1(\gamma) = \prod_{\bm{e}\in\gamma}W_{1}(\bm{e})$ is the product of Wilson links along an ordered path of edges $\gamma$ around the plaquette $\bm{p}$. $E_{\bm{e}}^2$ is the quadratic Casimir operator for the associated gauge group, which is diagonal in the irrep basis. The term $H_K$ contains the fermionic hopping terms which must be dressed by a Wilson link for gauge invariance. This also couples the matter and the gauge field with coupling strength $g_K$.

The associated Hilbert space is spanned by all possible configurations of fermionic and gauge fields. However, we are only interested in the physical subspace which corresponds to field configurations satisfying Gauss's law at each vertex in the lattice. In particular, Gauss's law annihilates nonphysical states according to
\al{
    \sum_{\bm{e}\in\delta\bm{v}}G(\bm{e})|\psi\rangle=0
}
where $G(\bm{e})$ is the Gauss operator. All of the physical operators must also commute with the Gauss's law operator, which ensures that states remain in the physical subspace.

\subsection{$q$-Deformed Pure KS Hamiltonian}
\label{sec:Deformed-YM-Theory}
Prior work on $q$-deformed KS theory has focused on the pure theory, \emph{i.e.}, with no matter~\cite{Zache:2023b,Hayata:2023a,hayata2023string}.
Herein, the regularization of the pure KS theory involves replacing the the Lie group $G$ with a quantum group $G_k$ constructed by a deformation of the universal enveloping algebra at a root of unity controlled by $k$.
While $G$ contains a countably infinite number of irreps which satisfy rules according to the Lie algebra $\mathfrak{g}$, $G_k$ contains a finite number of irreps controlled by the deformation parameter $k$, and which satisfy a \emph{fusion algebra} as in Eq.~\eqref{eq:fusion}.
The fusion coefficients $N^c_{ab}$ are derived from the deformed universal enveloping algebra.
The quantum group $G_k$ limits to the undeformed Lie group $G$ as $k\rightarrow \infty$.

\begin{figure}
    \includegraphics[width=\linewidth]{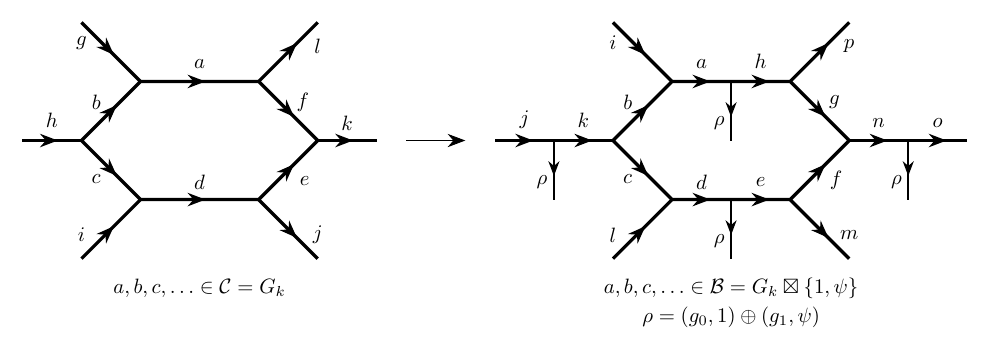}
    \caption{\label{fig:pure_gauge_to_matter} Starting from the deformed pure-gauge theory for gauge group $G_k$, defined on a 2d trivalent lattice (left), matter is introduced by the inclusion of additional dangling edges on horizontal edges in a 2+1d lattice (right) which support physical charges on a composite anyon model $G_k\boxtimes \{1,\psi\}$. The dangling edges are assigned a fixed semisimple object in $\mathcal{B}$, which we take to be the direct sum of the vacuum and generating Wilson lines of the individual layers.}
\end{figure}

As the fusion product is naturally trivalent, the lattice upon which the $q$-deformed Hamiltonian is defined must be resolved into a trivalent lattice as well. 
This is accomplished by point-splitting the vertices of the square lattice hosting the usual KS Hamiltonian, which yields a hexagonal lattice (see left side of Fig.~\ref{fig:pure_gauge_to_matter}). 
The $q$-deformed Hilbert space then consists of assigning each edge of the hexagonal lattice with an irrep from $G_k$, with the Gauss law at each trivalent vertex being satisfied if the fusion product is satisfied at the vertex.
This formulation reproduces the spin-network representation of the pure KS Hamiltonian as $k\rightarrow \infty$, which is equivalent to the quadrivalent KS formulation upon applying the Gauss' law constraint~\cite{Zache:2023b}.

The $q$-deformed Hamiltonian consists of an electric field term acting on the edges and a magnetic term on plaquettes.
The electric field term is given by the $q$-deformed quadratic Casimir operator of the gauge group
\al{
    E_{\bm{e}}^2 \ket{j}_{\bm{e}} = [C(j)]_q \ket{j}_{\bm{e}},
}
where $[n]_q \equiv (1-q^n)/(1-q)$ is a $q$-number for an integer $n$. 
The matrix elements of the plaquette operator are computed from the $F$ symbols of the fusion model, a proof for which can be found in the Supplementary Material of Ref.~\cite{Zache:2023b}:
\al{
\begin{aligned}
T_{\bm{p}}^\alpha 
\sbox0{\deformedlattice}
    \mathopen{\resizebox{1.2\width}{\ht0}{$\Biggl |$}}
    \usebox{0}
    \mathclose{\resizebox{1.2\width}{\ht0}{$\Biggr\rangle$}}
    &=\sbox0{\deformedlatticeloop}
    \mathopen{\resizebox{1.2\width}{\ht0}{$\Biggl |$}}
    \usebox{0}
    \mathclose{\resizebox{1.2\width}{\ht0}{$\Biggr\rangle$}} \\
    &=\sum_{a',b',\ldots,f'}
    [F_{i}^{a' \alpha \overline{b}}]_{a\overline{b}'}
    [F_{h}^{b' \alpha c}]_{bc'}
    [F_{i}^{\overline{c}' \alpha d}]_{\overline{c}d'}
    [F_{\overline{j}}^{\overline{d}' \alpha e}]_{\overline{d}e'} \\
    &\qquad\qquad\qquad\times
    [F_{\overline{k}}^{\overline{e}' \alpha \overline{f}}]_{\overline{e}\:\overline{f}'}
    [F_{\overline{l}}^{f' \alpha \overline{a}}]_{f\overline{a}'}
    \sbox0{\deformedlatticeresolved}
    \mathopen{\resizebox{1.2\width}{\ht0}{$\Biggl |$}}
    \usebox{0}
    \mathclose{\resizebox{1.2\width}{\ht0}{$\Biggr\rangle$}}
\end{aligned}.
}
This form the plaquette operator was first established in the context of Levin-Wen models~\cite{levin2005string}.

The $q$-deformed regularization has a number of features that make it attractive for use in LGT simulations, particularly on a quantum computer. 
First, the regularization comes with an integer control parameter $k$ that precisely controls how large we take our local Hilbert spaces to be. 
This should be contrasted with the approximations of non-Abelian Lie groups by discrete subgroups~\cite{Lamm:2019a, Alexandru:2019a,Ji:2020a,Ji:2023a,Assi:2024a} where there is a largest discrete subgroup beyond which one is not able to improve the approximation. 
Second, we retain a notion of gauge invariance, as we constrain the edge labels to obey the fusion rules of the category, implementing a version of Gauss's law.
Third, as we will show later, compiling the quantum circuits necessary to implement the algorithmic primitives of modern quantum simulation algorithms is quite natural: the electric field strength term is diagonal in the electric basis, and the magnetic term is expressed explicitly in terms of $F$ symbols, which are tensors that can be implemented in terms of unitary matrices.

Currently, the $q$-deformed lattice gauge theory is only formulated for the case of pure gauge theory for (semi-)simple gauge groups, with a proposal for coupling to hard-core bosonic matter given in Ref.~\cite{Zache:2023b}. 
In order to be potentially useful to particle physics, a formulation which couples the gauge fields to fermionic matter is necessary. 
It is also prudent to consider a generalization which permits non-semi-simple gauge groups of interest, like $U(N)$.
In the next section, we present such a framework using \emph{fusion surface models} and provide explicit quantum circuit primitive constructions for realizing quantum simulation of anyonic-regularized LGTs coupled to fermionic matter.

%%%%%%%%%%%%%%%%%%%%%%%%%%%%%%%%%%%%%%%%%%%%%%%%%%%%%%%%%%%%%%%%%%%%

\section{Anyonic Regularization}
\label{sec:Deformed_KS_Hamiltonian}

%%%%%%%%%%%%%%%%%%%%%%%%%%%%%%%%%%%%%%%%%%%%%%%%%%%%%%%%%%%%%%%%%%%%

Coupling the anyonic-regularized gauge theory to electrically charged fermionic matter will require us to expand our Hilbert space to include states with charged particles. From the point of view of the KS Hamiltonian, it would be natural to place fermionic modes at the vertices of the lattice and to enforce Gauss's law saying that the net electric flux out of a vertex should be equal to the charge at the vertex. One issue with this is that it would complicate the implementation of the plaquette operator which is defined in terms of $F$ and $R$ symbols. Since these symbols are only defined when the vertices obey the fusion rules, allowing for violations of the fusion rules would lead to an ambiguity as to how to proceed with defining the plaquette operator.

After regularizing the gauge theory, the Wilson lines become the worldlines of anyons of a corresponding Chern-Simons theory. The Hilbert space of our gauge-matter system will therefore have the interpretation as the state space of a system of interacting anyons. Models with precisely this interpretation are known in the literature as \textit{anyon chains} in 1d~\cite{Feiguin:2007a} and as \textit{fusion surface models} in 2d~\cite{Inamura:2024a} and we will utilize the technology developed in these papers to construct the state space of our system and some of the operators. Note that other recent works have considered interpreting the anyonic content of a topological phase as a kind of charged matter in a gauge theory in a different context~\cite{zhao2024noninvertible,zhao2025nonabelian,zhao2025landau}.

Our state space will consist of labelings of the edges of a trivalent graph by the objects of a category, as in the case of the $q$-deformed pure KS theory.
We will, however, attach additional ``dangling" edges labeled by a fixed anyon $\rho$, as in Fig.~\ref{fig:pure_gauge_to_matter}.
The introduction of dangling edges requires including braiding operations into our model, and hence we must generalize from an ordinary fusion algebra to a braided fusion category.

The braided fusion category we will take as input to our construction will be the product of $G_k$, the anyon model of the Chern-Simons theory for the gauge group $G$ at level $k$, and the braided fusion category $\{1,\psi\}$ consisting of just the vacuum and a fermion,
\al{
    \mathcal{B} = G_k \boxtimes \{1,\psi\}.
}
We can think of this as taking a ``stack" of the two anyon models. The objects of this product category are then pairs of objects from each layer with the first layer carrying the charges and the second layer providing fermionic statistics. 
It should be noted that in the case of a (semi-)simple gauge group $G$, the Chern-Simons anyon model has identical fusion rules to the $q$-deformed quantum group, while also containing braiding rules not found in the quantum group formulation; in the case of non-semi-simple gauge groups, the quantum group construction is inapplicable entirely.

We will bind the charges to the fermionic particles by taking the anyon $\rho$ to be the following semi-simple object
\al{
    \rho = (g_0,1)\oplus(g_1,\psi)
}
where $g_0$ is the vacuum particle corresponding to the trivial representation of the gauge group, $g_1$ is the anyon corresponding to the fundamental representation, $1$ is the absence of a fermion, and $\psi$ is the fermion. Taking $\rho$ to be the semisimple object $\rho =(g_0,1)\oplus(g_1,\psi)$ means that at each dangling edge, we can have either $(g_0,1)$, no charged particle, or $(g_1,\psi)$, a charged particle. The dynamical degrees of freedom live on the non-dangling edges of the lattice and have basis states labeled by the objects of $\mathcal{B}$. The fusion rules of $\mathcal{B}$ are then enforced at every vertex. 

Note that this formulation obscures the internal degrees of freedom of the representations carried by charged particles. We interpret this as a feature of this formulation, rather than a bug, as we are minimally encoding the gauge invariant states and operators of the system. Facets that are subject to gauge transformation are not locally accessible and should be interpreted as being dissolved into the fusion spaces of the anyon model.

\subsection{Fermionic Hamiltonians}
\label{sec:fermionic_Hamiltonians}

As a warm-up to coupling the anyonic-regularized gauge theory to fermionic matter, we will demonstrate the construction of a purely fermionic Hamiltonian within the fusion surface model framework. This will hopefully also serve as a demonstration of how a spin Hamiltonian can be obtained from the more abstract diagrammatic operators to come. We will show explicitly how generators for the even-parity subalgebra of fermionic operators can be constructed in terms of the diagrammatic operators presented in Sec.~\ref{sec:Constructing_Hamiltonian}. We will find that the obtained operators reproduce those of the Bravyi-Kitaev superfast encoding/exact bosonization~\cite{bravyi2002fermionic,chen2018exact}.

Recall that the fermionic operators $a_x^\dagger$ and $a_x^\pd$ create and annihilate a fermion at site $x$, respectively, and satisfy the anticommutation relations
\al{
    \{a_x^\dagger,a_y^\dagger\}=\{a_x^\pd,a_y^\pd\}=0\,, \quad \{a_x^\pd,a_y^\dagger\}=\delta_{xy}\mathbb{I}\,.
}
Fermionic Hamiltonians in general are represented via sums of even-parity products of these operators. Alternatively, one can define Majorana operators
\al{
    \gamma_x^\po=a_x^\pd+a_x^\dagger\,,\quad \overline{\gamma}_x=-i(a_x^\pd-a_x^\dagger)\,,
}
which satisfy the anticommutation relations
\al{
    \{\gamma_x^\po,\gamma_y^\po\}=\{\overline{\gamma}_x,\overline{\gamma}_y\}=2\delta_{xy}\mathbb{I}\,,\quad \{\gamma_x^\po,\overline{\gamma}_y\}=0\,.
}
A useful (overcomplete) generating set of even-parity fermionic operators consists of the following operators along edges, $\bm{e}$, of the lattice
\al{
    O_{\bm{e}}^\psi=-i\overline{\gamma}_{\partial\bm{e}_0}\gamma_{\partial\bm{e}_1}^\po\,,
}
which flip the fermion parity of the modes at the endpoints of the edge $\bm{e}$ and the following single-site fermion parity operator associated to the vertex $\bm{v}$
\al{
    P_{\bm{v}}^{\psi} = -i\gamma_{\bm{v}}^{\po}  \overline{\gamma}_{\bm{v}}
}

Since we are focused solely on the fermionic algebra here, we take as input the braided fusion category of supervector spaces~\cite{aasen2019fermion}, $\mathcal{B}=\{1,\psi\}$, consisting only of the vacuum particle and a fermion. The anyons obey $\mathbb{Z}_2$ fusion rules, $\psi \times \psi = 1$, and the $F$ symbol is trivial, with all elements being just zero or one as allowed by the fusion rules. The $R$ symbol has a single nontrivial element, $R^{\psi \psi}_{1} = -1$, which provides the fermionic statistics of $\psi$. Correspondingly, $\psi$ has a topological spin of $\theta_\psi=-1$. 

Since the braided fusion category we use has two simple objects, each edge of our lattice will carry a single qubit. Due to the triviality of the $F$ symbols and the Abelian nature of the fusion rules, we will also neglect the trivalent resolution for simplicity, so the lattice we consider for this warm-up consists of a square lattice with dangling edges at vertices. Since we want to allow each vertex to be either occupied by a fermion or unoccupied, we take the dangling edge to be labeled by the semi-simple object $\rho=1\oplus\psi$. The $1$ state is interpreted as the empty mode, and the $\psi$ state is the occupied mode. Once the state of the four edges incident to a given vertex is specified, the uniqueness of the fusion channels fixes the fermion parity of the vertex.

\begin{figure}[t]
    \centering
    \includegraphics[width=0.6\linewidth]{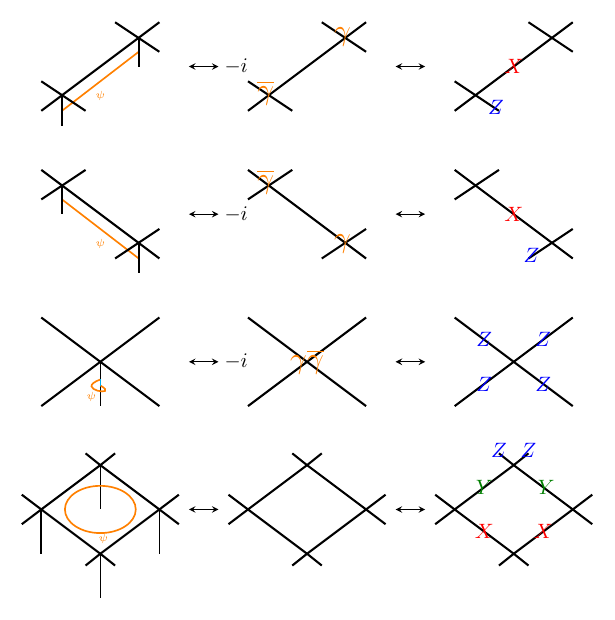}
    \caption{The fermionic operators generating the even parity subalgebra (top two rows) Majorana bilinears flipping the parity of the modes at the endpoints of the edge (third row) Single-site parity operator (bottom row) Minimal 1-form symmetry loop acting trivially on the encoded fermionic system.}
    \label{fig:fermion_operators}
\end{figure}

Fermionic operators are defined by gluing on diagrams in the space $\Hom(\rho^{\otimes (L\times L)},\rho^{\otimes (L\times L)})$ to the dangling edges for an $L\times L$ lattice. One then uses the $F$ and $R$ symbols to absorb or remove the extra lines in the diagram, recovering the form of the original diagram that defines the Hilbert space. In doing so, the elements of the $F$ and $R$ symbols provide the matrix elements of the operator on the Hilbert space. As long as the fusion rules are obeyed at each vertex in the diagrammatic operator (which they must be) the operator will preserve the total fermion parity. A consequence of having all of the operators defined in terms of anyon diagrams glued in from the bottom is that any Hamiltonian we construct will obey the 1-form symmetry obtained by gluing in closed loops of the fermion line from above, as in the bottom of Fig.~\ref{fig:fermion_operators}. Having this 1-form symmetry with $\mathbb{Z}_2$ fusion rules and the nontrivial braiding phase is equivalent to the fact that the spin system we obtain has an emergent fermion.

We define the operators $O_{\bm{e}}^\psi$ by gluing on a $\psi$ string attached to two dangling edges as in Eqs.~\eqref{eq:hopping_operator_x} and~\eqref{eq:hopping_operator_y}. Using the $F$ and $R$ moves to resolve multiple such strings into the lattice, we can determine the commutation relations of these operators. With these commutation relations in hand, we can make the simple identification that if a $\psi$ line enters the dangling edge from the left side of the page, it corresponds to the Majorana operator $\gamma$ and if the $\psi$ line enters a dangling edge from the right side of the page, it corresponds to the Majorana operator $\overline{\gamma}$. This allows one to identify the hopping operator with the Majorana bilinear 
\al{
    O_{\bm{e}}^\psi=-i\overline{\gamma}_{\partial\bm{e}_0}\gamma_{\partial\bm{e}_1}^\po\,,
}
where $\partial\bm{e}_i$ are the boundaries, or vertices, of the associated edge $\bm{e}$. The single-site fermion parity operator can be realized by attaching a small $\psi$ line to one side of the dangling edge, crossing over, and attaching to the other side. The nontrivial sign in the $R$ symbol provides the $(-1)$ phase when $P_{\bm{v}}^\psi=-i\gamma^\po_{\bm{v}}\overline{\gamma}_{\bm{v}}$ acts on an occupied mode.

We then obtain all remaining Majorana bilinears using the parity operators in Eq.~\eqref{eq:parity_operator_rho} using the fact that $P_{\bm{v}}^\psi=-i\gamma^\po_{\bm{v}}\overline{\gamma}_{\bm{v}}$, $\gamma_{\bm{v}}^\po=iP_{\bm{v}}^{\psi}\overline{\gamma}_{\bm{v}}$, and $\overline{\gamma}_{\bm{v}}=-iP_{\bm{v}}^\psi\gamma_{\bm{v}}$: 
\al{
\begin{aligned}
    P_{\partial\bm{e}_0}^\psi O_{\bm{e}}^\psi &=-\gamma_{\partial\bm{e}_0}^\po \gamma_{\partial\bm{e}_1}^\po\,, \\
    P_{\partial\bm{e}_1}^\psi O_{\bm{e}}^\psi &=\overline{\gamma}_{\partial\bm{e}_0} \overline{\gamma}_{\partial\bm{e}_1}\,, \\
    P_{\partial\bm{e}_0}^\psi P_{\partial\bm{e}_1}^\psi O_{\bm{e}}^\psi &=-i\gamma_{\partial\bm{e}_0}^\po \overline{\gamma}_{\partial\bm{e}_1}\,.
    \label{eq:paritytobilinear}
\end{aligned}
}
Using these relations, we can generate the even parity subalgebra in terms of these diagrammatic parity and hopping operators. 

The fermion loop can be resolved into the lattice via
\al{\label{eq:fermion_loop}
    \begin{aligned}
    \sbox0{\fermionloopzero}
    \usebox{0}
    &=
    \sbox0{\fermionloopone}
    \usebox{0}
    =
    \sbox0{\fermionlooptwo}
    \usebox{0} \\
    &=\sbox0{\fermionloopthree}
    \usebox{0}
    =P_{\bm{v}}^\psi\quad
    \sbox0{\fermionloopfour}
    \usebox{0} \\
    &=(-1)^{\delta_{\psi,a\times b\times c\times d}}\quad
    \sbox0{\fermionloopfour}
    \usebox{0}
    =\sbox0{\fermionloopfive}
    \usebox{0}
\end{aligned}
}
All of the diagrammatic fermion operators commute with such operators. These operators therefore generate a symmetry. More generally, fusing in a closed $\psi$ line from above will commute with all the other even-parity operators in Fig.~\ref{fig:fermion_operators}. This is therefore a 1-form symmetry of codimension-1 (in space) topological defects which fuse according to the rules of the input braided fusion category.

To map the diagrammatic operators to operators on the system of qubits, we identify the $1$ label on an edge with the $|0\rangle$ state and the $\psi$ label with the $|1\rangle$ state and explicit operators on the multi-qubit Hilbert space will follow. When a $\psi$ line fuses into an edge, the $\mathbb{Z}_2$ fusion rules flip the label on the edge, which in the qubit language amounts to a Pauli-$X$ operator. The label on the dangling edge can be uniquely inferred by looking at the labels of the edges sharing the vertex. So the parity of the mode, can be obtained by acting with Pauli-$Z$ operators on the qubits on the edges incident to the vertex. Explicit qubit operators for the generating set of even-parity operators are given in Fig.~\ref{fig:fermion_operators}. 

The operators in the right column of Fig.~\ref{fig:fermion_operators} have appeared previously in the superfast encoding/exact bosonization~\cite{bravyi2002fermionic,chen2018exact}. We can thus give the encoding of the fermionic system the interpretation as having gauged the fermion parity. Gauging the global fermion parity is known to result in a system with a 1-form symmetry~\cite{levin2003fermions,gaiotto2016spin,chen2018exact,chen2020exact}, shown on the bottom row of Fig.~\ref{fig:fermion_operators}. In the fusion surface model framework that we employed, our starting point was the 1-form symmetry that we knew we would obtain through such a gauging. This illustrates the spirit of the formalism, the desired symmetry structure is a very useful starting point when engineering models with certain properties. Here in this warm-up, it was arbitrary fermionic Hamiltonians for which there are many ways of encoding~\cite{chien2026simulating}, but it was natural to use the common formulation. For the remainder of the paper, we will use the framework for regularizations of lattice gauge theories coupled to the fermionic matter we have discussed here.

Again, as noted above, we neglected to resolve the vertices into trivalent vertices in this simple case. In what comes, we will resolve the vertices. We expand the square lattice into a hexagonal lattice and place dangling edges on the horizontal edges only. We still have a square lattice of dangling edges and there will be a one-to-one correspondence of plaquettes between the resolved and this unresolved lattice.

%%%%%%%%%%%%%%%%%%%%%%%%%%%%%%%%%%%%%%%%%%%%%%%%%%%%%%%%%%%%%%%%%%%%

\subsection{Anyonic-Regularized KS Hamiltonian}
\label{sec:Constructing_Hamiltonian}

We now generalize our construction for fermionic Hamiltonians to that of the full KS Hamiltonian. Taking as a starting point the Kogut-Sussind Hamiltonian of Eq.~\eqref{eq:KS_Hamiltonian}, we will demonstrate how to construct each of these interactions in terms of the $F$ and $R$ symbols of the input braided fusion category $\mathcal{B}$.

Starting with the magnetic term, we define the plaquette operator $T_{\bm{p}}^\alpha$, which is defined in exactly the same way as in Refs.~\cite{Zache:2023b, Hayata:2023a}, by inserting a Wilson loop of charge $\alpha\in\mathcal{B}$ around a plaquette in the 2d plane. However, as a result of the dangling edges in our construction, the anyon crosses over a dangling edge, meaning that the decomposition of $T_{\bm{p}}^\alpha$ will now contain a product of both $F$ and $R$ symbols in contrast to prior results for pure-gauge theories. The plaquette operator and its decomposition are given explicitly by:
\al{\label{eq:plaquette_operator}
    \begin{aligned}
    T_{\bm{p}}^\alpha
    \sbox0{\lattice}
    \mathopen{\resizebox{1.2\width}{\ht0}{$\Biggl |$}}
    \usebox{0}
    \mathclose{\resizebox{1.2\width}{\ht0}{$\Biggr\rangle$}}
    &=
    \sbox0{\latticeplaquette}
    \mathopen{\resizebox{1.2\width}{\ht0}{$\Biggl |$}}
    \usebox{0}
    \mathclose{\resizebox{1.2\width}{\ht0}{$\Biggr\rangle$}} \\
    &=\sum_{a',b',\ldots,h'}
    [F_{i}^{a' \alpha \overline{b}}]_{a\overline{b}'}
    [F_{k}^{b' \alpha c}]_{bc'}
    [F_{l}^{\overline{c}' \alpha d}]_{\overline{c}d'}
    [F_{\overline{\rho}}^{\overline{d}' \alpha e}]_{\overline{d}e'}
    [F_{\overline{m}}^{\overline{e}' \alpha f}]_{\overline{e}f'}
    [F_{\overline{n}}^{\overline{f}' \alpha \overline{g}}]_{\overline{f}\:\overline{g}'} \\
    &\quad\:\:\:\times [F_{\overline{p}}^{g' \alpha \overline{h}}]_{g\overline{h}'}
    [F_{\overline{h}}^{\rho' \alpha \overline{a}}]_{\rho \overline{a}'}
    [F_{a'}^{h' \alpha \rho'}]_{h\rho}
    (R_{\rho}^{\overline{\alpha}\rho'})^{-1}
    \sbox0{\latticeplaquettefused}
    \mathopen{\resizebox{1.2\width}{\ht0}{$\Biggl |$}}
    \usebox{0}
    \mathclose{\resizebox{1.2\width}{\ht0}{$\Biggr\rangle$}}\,,
\end{aligned}
}
where the sum runs over the admissible fusion products resulting from resolving the Wilson loop into the lattice, \emph{e.g.}, $a\times\alpha=\sum_{a'}a'$ for all valid fusion products.

Next, we consider the electric term. Recall that the Casimir operator measures the strength of the electric field along an edge $\bm{e}$, and thus acts nontrivially only on the gauge group layer $G_k$ of the full braided fusion category $\mathcal{B}$. Since we are working in the irrep basis, this operator is diagonal and we can start by defining a loop operator 
\al{
L_{\bm{e}}^c
    \sbox0{\latticeelectric}
    \mathopen{\resizebox{1.2\width}{\ht0}{$\Biggl |$}}
    \usebox{0}
    \mathclose{\resizebox{1.2\width}{\ht0}{$\Biggr\rangle$}}
    =
    \sbox0{\latticeelectricloop}
    \mathopen{\resizebox{1.2\width}{\ht0}{$\Biggl |$}}
    \usebox{0}
    \mathclose{\resizebox{1.2\width}{\ht0}{$\Biggr\rangle$}}
}
which wraps a Wilson loop $c$ around a lattice edge carrying an anyon $b$. In terms of $L_{\bm{e}}^c$, we can then define a projector onto an anyon $a$ via a sum, weighted by the modular $S$ matrix, over all the anyons $c\in G_k$ as follows:
\al{
\begin{aligned}
    \Pi_a
    \sbox0{\latticeelectric}
    \mathopen{\resizebox{1.2\width}{\ht0}{$\Biggl |$}}
    \usebox{0}
    \mathclose{\resizebox{1.2\width}{\ht0}{$\Biggr\rangle$}}&=\sum_{c\in G_k} S_{0a}\overline{S}_{ca}
    \sbox0{\latticeelectricloop}
    \mathopen{\resizebox{1.2\width}{\ht0}{$\Biggl |$}}
    \usebox{0}
    \mathclose{\resizebox{1.2\width}{\ht0}{$\Biggr\rangle$}}
    =\sum_{c\in G_k}S_{0a}\overline{S}_{ca}\frac{S_{cb}}{S_{0b}}
    \sbox0{\latticeelectric}
    \mathopen{\resizebox{1.2\width}{\ht0}{$\Biggl |$}}
    \usebox{0}
    \mathclose{\resizebox{1.2\width}{\ht0}{$\Biggr\rangle$}}
    =\frac{S_{0a}}{S_{0b}}\sum_{c\in G_k} \overline{S}_{ac}S_{cb}
    \sbox0{\latticeelectric}
    \mathopen{\resizebox{1.2\width}{\ht0}{$\Biggl |$}}
    \usebox{0}
    \mathclose{\resizebox{1.2\width}{\ht0}{$\Biggr\rangle$}} \\
    &=\frac{S_{0a}}{S_{0b}}(S^\dagger S)_{ab}
    \sbox0{\latticeelectric}
    \mathopen{\resizebox{1.2\width}{\ht0}{$\Biggl |$}}
    \usebox{0}
    \mathclose{\resizebox{1.2\width}{\ht0}{$\Biggr\rangle$}} 
    =\delta_{ab}
    \sbox0{\latticeelectric}
    \mathopen{\resizebox{1.2\width}{\ht0}{$\Biggl |$}}
    \usebox{0}
    \mathclose{\resizebox{1.2\width}{\ht0}{$\Biggr\rangle$}}\,.
\end{aligned}
}
Using this projector, we can then construct the quadratic Casimir operator via
\al{
\begin{aligned}
    E_{\bm{e}}^2&=\sum_{a\in G_k}\epsilon(a)\Pi_a
    =\sum_{a,b\in G_k}\varepsilon(a) S_{0a}\overline{S}_{ba}L_{\bm{e}}^b=\sum_{b\in G_k}\beta_b L_{\bm{e}}^b
    \label{eq:electric_field}
\end{aligned}
}
where we project onto each anyon worldline $a$ with strength $\epsilon(a)$, which corresponds to the $q$-deformed electric field strength. These electric operators can be seen as explicitly breaking the 1-form symmetries of the fusion surface model~\cite{gaiotto2015generalized,Inamura:2024a}.

We now turn to the Hamiltonian terms containing interactions with fermionic matter, for which we present two new operators.
First considering the mass term, we introduce the parity operator $P_{\bm{v}}^\psi$, which attaches a Wilson line $\psi$ to a dangling edge, defined by
\al{\label{eq:parity_operator_rho}
    P_{\bm{v}}^\psi\sbox0{\latticemass}
    \mathopen{\resizebox{1.2\width}{\ht0}{$\Biggl |$}}
    \usebox{0}
    \mathclose{\resizebox{1.2\width}{\ht0}{$\Biggr\rangle$}}
    =
    \sbox0{\latticemasstwist}
    \mathopen{\resizebox{1.2\width}{\ht0}{$\Biggl |$}}
    \usebox{0}
    \mathclose{\resizebox{1.2\width}{\ht0}{$\Biggr\rangle$}}
    = (-1)^{\rho}
    \sbox0{\latticemass}
    \mathopen{\resizebox{1.2\width}{\ht0}{$\Biggl |$}}
    \usebox{0}
    \mathclose{\resizebox{1.2\width}{\ht0}{$\Biggr\rangle$}}\,,
}
which acquires a phase depending on the charge label $\rho\in\mathcal{B}$. This distinguishes between the two anyons $(g_0,1)$ and $(g_1,\psi)$ which can be hosted on the dangling edge to indicate the presence of a fermion or not.

Lastly, we consider the Hamiltonian term coupling the fermionic matter and gauge fields, and define hopping operators $O_{\bm{e}}^\alpha$ which introduce a Wilson line $\alpha\in\mathcal{B}$ connecting adjacent dangling edges and facilitates topological charge transportation. Since our theory is formulated on a 2+1d lattice, there are two hopping operators defined by
\al{
    \begin{aligned}\label{eq:hopping_operator_x}
    O_{\bm{e},x}^\alpha
    \sbox0{\xlattice}
    \mathopen{\resizebox{1.2\width}{\ht0}{$\Biggl |$}}
    \usebox{0}
    \mathclose{\resizebox{1.2\width}{\ht0}{$\Biggr\rangle$}}
    &=
    \sbox0{\xlatticehopping}
    \mathopen{\resizebox{1.2\width}{\ht0}{$\Biggl |$}}
    \usebox{0}
    \mathclose{\resizebox{1.2\width}{\ht0}{$\Biggr\rangle$}} \\
    &=\sqrt{\frac{\text{d}_\rho\text{d}_\alpha}{\text{d}_{\rho'}}}\sum_{k',c',d'}
    [F_{j}^{\rho' \alpha k}]_{\rho k'}
    [F_{\overline{b}}^{\overline{k}' \alpha c}]_{\overline{k}c'}
    [F_{\overline{e}}^{\overline{d}' \alpha \rho}]_{\overline{d}\rho'} \\
    &\!\qquad\qquad\qquad\times
    \sum_{l'}\frac{\text{d}_{l'}}{\text{d}_l}
    [F_{c}^{\overline{l}' \alpha d}]_{\overline{l}d'} 
    [F_{\overline{d}'}^{\overline{l} \:\overline{\alpha}\: \overline{c}}]_{\overline{l}'\overline{c}'}   
    (R_{\overline{l}'}^{\overline{l} \:\overline{\alpha}})^{-1}
    \sbox0{\xlatticehoppingfused}
    \mathopen{\resizebox{1.2\width}{\ht0}{$\Biggl |$}}
    \usebox{0}
    \mathclose{\resizebox{1.2\width}{\ht0}{$\Biggr\rangle$}}\,,
    \end{aligned} \\
    \begin{aligned}\label{eq:hopping_operator_y}
    O_{\bm{e},y}^\alpha
    \sbox0{\ylattice}
    \mathopen{\resizebox{1.2\width}{\ht0}{$\Biggl |$}}
    \usebox{0}
    \mathclose{\resizebox{1.2\width}{\ht0}{$\Biggr\rangle$}}
    &=
    \sbox0{\ylatticehopping}
    \mathopen{\resizebox{1.2\width}{\ht0}{$\Biggl |$}}
    \usebox{0}
    \mathclose{\resizebox{1.2\width}{\ht0}{$\Biggr\rangle$}} \\
    &=\sqrt{\frac{\text{d}_\rho\text{d}_\alpha}{\text{d}_{\rho'}}}\sum_{k',b',a'}
    [F_{j}^{\rho' \alpha k}]_{\rho k'}
    [F_{i}^{\overline{b}' \alpha a}]_{\overline{b}a'}
    [F_{\overline{h}}^{\overline{a}' \alpha \rho}]_{\overline{a}\rho'}\\
    &\!\qquad\qquad\qquad\times
    \sum_{c'}\frac{\text{d}_{c'}}{\text{d}_c}
    [F_{k}^{c' \alpha b}]_{cb'}
    [F_{\overline{b}'}^{c \alpha \overline{k}}]_{\overline{k}'c'}
    (R_{c'}^{c\overline{\alpha}})^{-1}
    \sbox0{\ylatticehoppingfused}
    \mathopen{\resizebox{1.2\width}{\ht0}{$\Biggl |$}}
    \usebox{0}
    \mathclose{\resizebox{1.2\width}{\ht0}{$\Biggr\rangle$}}\,,
    \end{aligned}
}
corresponding to the two directions the fermions can hop on the lattice.

Using these diagrammatic operators, we can now fully express the KS Hamiltonian purely in terms of $F$ and $R$ symbols of the anyon model which naturally evokes an LCU decomposition. In particular, we have
\al{
    &\! H_M=\sum_{\bm{v}} (-1)^{\bm{v}} P_{\bm{v}}^\psi\,, \label{eq:Hamiltonian_mass_fusion} \\
    &\begin{aligned}\label{eq:Hamiltonian_kinetic_fusion}
    H_K&=
    \frac{1}{4}\sum_{\bm{e}}\gamma_{\partial\bm{e}_0}^\po\gamma_{\partial\bm{e}_1}^\po\left(O_{\bm{e}}^{\alpha\pd}-O_{\bm{e}}^{\alpha\dagger}\right)+i\gamma_{\partial\bm{e}_0}^\po\overline{\gamma}_{\partial\bm{e}_1}\left(O_{\bm{e}}^{\alpha\pd}+O_{\bm{e}}^{\alpha\dagger}\right) \\
    &\qquad\qquad -i\overline{\gamma}_{\partial\bm{e}_0}\gamma_{\partial\bm{e}_1}^\po\left(O_{\bm{e}}^{\alpha\pd}+O_{\bm{e}}^{\alpha\dagger}\right)+\overline{\gamma}_{\partial\bm{e}_0}\overline{\gamma}_{\partial\bm{e}_1}\left(O_{\bm{e}}^{\alpha\pd}-O_{\bm{e}}^{\alpha\dagger}\right)\,,
    \end{aligned} \\
    &\begin{aligned}\!\!\!\!\label{eq:Hamiltonian_yang_mills_fusion}
     H_{YM}
    &=\frac{g^2}{2}\sum_{\bm{e}}E_{\bm{e}}^2-\frac{1}{2a^2g^2}\sum_{\bm{p}}\left(T_{\bm{p}}^\alpha+T_{\bm{p}}^{\alpha\dagger}\right)\,,
    \end{aligned}
}
where we take $\alpha=(g_1,\psi)$ to be the charge carrier. Note that the trace on the Wilson loops is no longer present as they no longer correspond to taking a trace in a representation. The dimensions of the representations are replaced by quantum dimensions of the $\alpha$ anyons which may not even be integer-valued.

\section{Quantum Simulation of Anyonic-Regularized KS Hamiltonians}
\label{sec:Quantum_simulation}

With the general formulation of the KS Hamiltonian in terms of our diagrammatic operators, we will now discuss the details of simulating this Hamiltonian on a quantum computer. Here we consider explicit implementations for $U(1)_k$ and $SU(2)_k$ LGTs, leaving further generalizations to future work. Since the Hamiltonian of Eqs.~\eqref{eq:Hamiltonian_mass_fusion}--\eqref{eq:Hamiltonian_yang_mills_fusion} can be written as a linear combination of unitary operators using the $F$ and $R$ symbols, we can take advantage of this structure to realize any state-of-the-art eigenstate preparation or time-dynamics quantum simulation algorithm.

As an example we can consider the time dynamics algorithm of Ref.~\cite{Tong:2022a} which yields asymptotically near-optimal scaling in all relevant system parameters, and demonstrated remarkable improvements in explicit resource estimates for $U(1)$, $SU(2)$, and $SU(3)$ LGTs~\cite{Rhodes:2024a}. In this algorithm, one works in the interaction picture, for which the time-dependent Hamiltonian is
\al{
    H_I(t)=e^{itH_E}(H_M+H_K+H_B)e^{-itH_E}\,,
}
and we have explicitly decomposed the Yang-Mills Hamiltonian $H_{YM}$ into its electric, $H_E$, and magnetic, $H_B$, terms.

Since the electric term, corresponding to the quadratic Casimir operator, is diagonal in the electric basis, its time evolution can be fast-forwarded~\cite{Gu:2021a}. This only requires constructing unitary operators to prepare the eigenstates of the quadratic Casimir operators, namely, $[a^2]_q$ for $a\in U(1)_k$  and $[j(j+1)]_q$ for $j\in SU(2)_k$, and a series of phase gates~\cite{Childs:2004a}.

The remaining three terms can then be block encoded using the LCU oracles \textsc{prepare} and \textsc{select}~\cite{Low:2019a}. Since each of the Hamiltonian terms is written in the form $H=\sum_\ell \alpha_\ell U_\ell$, it can be encoded via queries to the oracles
\al{
    \textsc{prepare}|0\rangle &=\sum_\ell \sqrt{\frac{\alpha_\ell}{\|\alpha\|_1}}|\ell\rangle\,, \\
    \textsc{select}|\ell\rangle |\phi\rangle &=|\ell\rangle U_\ell |\phi\rangle\,,
}
where the number of Hamiltonian terms $\ell$ will be functions of the number of vertices, edges, and plaquettes in the lattice. The \textsc{prepare} oracle generates a superposition of index states, weighted by the coefficients of the Hamiltonian and the \textsc{select} oracle applies the unitary Hamiltonian terms $U_\ell$ to the basis states $|\phi\rangle$, controlled on the index states $|\ell\rangle$.

In order to implement \textsc{prepare}, we can make use of the techniques developed in Ref.~\cite{Rhodes:2024a} and absorb the nontrivial phases of the Hamiltonian terms into the unitary $F$ and $R$ symbols, such that $\textsc{prepare}$ amounts to generating uniform superpositions of index states. Then the action of \textsc{select} is simply to apply singly-controlled $F$ and $R$ symbols using the circuit primitives we will construct, which makes use of the unary iteration subroutine for indexing the states~\cite{Babbush:2018a}.

Upon successfully block encoding the Hamiltonian by querying these oracles, one can then simulate time dynamics via the truncated Dyson series techniques for time-dependent Hamiltonian simulation~\cite{Low:2019b, Berry:2020a}.

As we will see, the structure of the circuit primitives for the $F$ and $R$ symbols makes the post-Trotter simulation algorithm we sketched more enticing. However, we remark that any state-of-the-art Trotter simulation could also be applied using the circuit realizations of the $F$ and $R$ symbols. Thus, by providing only the circuit primitives, we open the door to detailed resource estimates and comparisons of various dynamical simulation algorithms, which we leave to future work. It only remains to show how to construct the quantum circuits realizing the $F$ and $R$ symbols of the input anyon model. We now provide these constructions for $U(1)_2$, $U(1)_k$, and $SU(2)_k$ LGTs.

%%%%%%%%%%%%%%%%%%%%%%%%%%%%%%%%%%%%%%%%%%%%%%%%%%%%%%%%%%%%%%%%%%%%
%%%%%%%%%%%%%%%%%%%%%%%%%%%%%%%%%%%%%%%%%%%%%%%%%%%%%%%%%%%%%%%%%%%%

\subsubsection{$U(1)_2$}
\label{sec:U(1)_2}

%%%%%%%%%%%%%%%%%%%%%%%%%%%%%%%%%%%%%%%%%%%%%%%%%%%%%%%%%%%%%%%%%%%%

The $U(1)_2$ anyon model is an Abelian gauge theory describing the semion edge theory of topological phases. It is equivalent to $SU(2)_1$ and $\mathbb{Z}_{2}^{(1/2)}$ braided fusion categories. The input braided fusion category is
\al{
    \mathcal{B} = U(1)_2 \boxtimes \{1,\psi\}\,,
}
where the topological excitations of $U(1)_2$ are $\{1,s\}$, consisting of the vacuum and a semion. Thus, the composite system contains four anyons $\{(1,1),(s,1),(1,\psi),(s,\psi)\}$, each of which is self-dual. This model thus obeys $\mathbb{Z}_2^{(1/2)}\times\mathbb{Z}_2$ fusion rules. The topological spins of the composite system are given by the product of topological spins of the individual layers. Namely, $\theta_{(1,1)}=1$, $\theta_{(s,1)}=i$, $\theta_{(1,\psi)}=-1$, and $\theta_{(s,\psi)}=-i$. On the dangling edges, we take the fixed lines $\rho$ to be
\al{
    \rho=(1,1)\oplus(s,\psi)\,.
}

In this model, the Hamiltonian terms of Eqs.~\eqref{eq:Hamiltonian_mass_fusion}-\eqref{eq:Hamiltonian_yang_mills_fusion} become
\al{
    &H_M=\sum_{\bm{v}} P_{\bm{v}}^\psi = \sum_{\bm{v}}(-1)^{\rho_{\bm{v}}}\,, \\
    &\begin{aligned}
    H_K&=\frac{1}{2}\sum_{\bm{e}}(i\gamma_{\partial\bm{e}_0}^\po\overline{\gamma}_{\partial\bm{e}_1}-i\overline{\gamma}_{\partial\bm{e}_0}\gamma_{\partial\bm{e}_1}^\po)O_{\bm{e}}^{(s,1)} \\
    &=\frac{1}{2}\sum_{\bm{e}}(\mathbb{I}-P_{\partial\bm{e}_0}^\psi P_{\partial\bm{e}_1}^\psi)O_{\bm{e}}^{(s,\psi)} \\
    &=\sum_{\bm{e}}\delta_{\partial\bm{e}_0,\partial\bm{e}_1+1}O_{\bm{e}}^{(s,\psi)}\,,
    \end{aligned} \\
    &\!\!\!\!H_{YM}=\frac{g^2}{2}\sum_{\bm{e}}E_{\bm{e}}^2-\frac{1}{a^2g^2}\sum_{\bm{p}}T_{\bm{p}}^{(s,\psi)}\,,
}
where for $H_K$ we have used the fact that since the anyons are self-dual then $O_{\bm{e}}^{\alpha\pd}=O_{\bm{e}}^{\alpha\dagger}$. Thus, $O_{\bm{e}}^\alpha$ only facilitates hopping when one vertex is occupied by a fermion and the other is not. For the electric term, the $q$-deformed electric field strength is $\epsilon(a)=[a^2]_q$ with $a\in\{1,s\}$.

In the $U(1)_2$ anyon model, each composite anyon can be stored in a two-qubit register with the first qubit storing the anyon from the $U(1)_2$ layer and the second qubit storing the anyon from the $\{1,\psi\}$ layer. Since $\{1,\psi\}$ obeys $\mathbb{Z}_2$ fusion rules, the nonzero $F$ symbols are all one, allowing us to consider only the action on the $U(1)_2$ layer for constructing the $F$ symbols. In particular, the $U(1)_2$ layer has $\mathbb{Z}_2^{(1/2)}$ fusion rules defined by
\al{
    [F^{abc}_d]_{e,f}^{\phantom{a}}=[F^{abc}_{a\oplus b\oplus c}]_{a\oplus b,b\oplus c}^{\phantom{a}}=e^{\frac{\pi i}{2}a(b+c-(b\oplus c))}\,,
}
where we have used $\oplus$ to indicate addition modulo two. Then the action on the $U(1)_2$ layer of our encoded states has the form
\al{
    |a,b;a\oplus b\rangle |a\oplus b,c;a\oplus b\oplus c\rangle\mapsto e^{\frac{\pi i}{2}a(b+c-(b \oplus c))} |a,b\oplus c;a\oplus b\oplus c\rangle |b,c;b\oplus c\rangle\,,
}
where encoded states store trivalent vertices, corresponding to their fusions. Note that an alternative encoding corresponds to storing the edges participating in an $F$ move, which is more space-efficient since it does not store repeated labels, but is less time-efficient.

\begin{figure}
    \centering
     \begin{subfigure}[t]{0.5\textwidth}
        \centering
            \includegraphics[width=\textwidth]{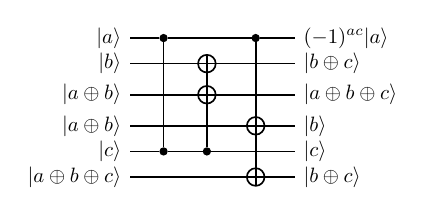}
            \caption{$F$ symbol}
            \label{fig:U(1)_2-Fmove}
     \end{subfigure}\hfill
    \begin{subfigure}[t]{0.5\textwidth}
        \centering
            \includegraphics[width=\textwidth]{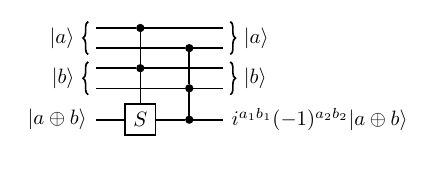}
            \caption{$R$ symbol}
            \label{fig:U(1)_2-Rmove}
     \end{subfigure}
    \caption{\label{fig:U(1)_2-circuits} Explicit circuit implementations for the primitive unitary operations of the $U(1)_2$ LGT Hamiltonian. The (a) $F$ symbol applies a phase if edges $a$ and $c$ are both semions, since $b$ is always a semion, and then changes the basis states via modular arithmetic. The (b) $R$ symbol applies a phase depending on the state of both the $U(1)_2$ layer and the $\{1,\psi\}$ layer, as indicated by label subscripts.}
\end{figure}

For this particular $F$ symbol, notice that $F=-1$ only when $a=b=c=1$, corresponding to all semions, and otherwise $F=1$. Moreover, we can take advantage of the structure of the $F$-\ symbols that appear in our Hamiltonian. Each $F$ move has the form
\al{
    [F^{abc}_d]_{e,f}=[F^{a'\alpha c}_d]_{a,c'}\,,
}
in which $\alpha=(s,\psi)$ always and primed variables denote fusion products with $\alpha$, \emph{e.g.}, $a'=a\times \alpha$. This allows for additional simplifications in the circuit construction using the fact that $b=\alpha=1$ always. 
This operation can be realized using the circuit given in Fig.~\ref{fig:U(1)_2-Fmove}.

Turning now to the $R$ symbols, there are contributions to the final phase from both layers. In particular, for the $\{1,\psi\}$ layer, $R^{\psi\psi}_1=-1$ while $R=1$ otherwise, and in the $U(1)_2$ layer $R^{ss}_1=i$, while $R=1$ otherwise. The action on the encoded state is then
\al{
    |a,b;a\oplus b\rangle \mapsto e^{\pi i(a_1 b_1/2 + a_2 b_2)}|a,b;a\oplus b\rangle\,,
}
where the subscripts index the particular layer, to which the phase depends on. This action can be realized via the circuit in Fig.~\ref{fig:U(1)_2-Rmove}.

%%%%%%%%%%%%%%%%%%%%%%%%%%%%%%%%%%%%%%%%%%%%%%%%%%%%%%%%%%%%%%%%%%%%

\subsubsection{$U(1)_k$}
\label{sec:U(1)_k}

%%%%%%%%%%%%%%%%%%%%%%%%%%%%%%%%%%%%%%%%%%%%%%%%%%%%%%%%%%%%%%%%%%%%

We now generalize from $U(1)_2$ to $U(1)_k$, for which the input braided fusion category becomes 
\al{
    \mathcal{B}=U(1)_k \boxtimes \{1,\psi\}\,,
}
where $U(1)_k$ is equivalent to the modular theory $\mathbb{Z}_k^{(1/2)}$ for even $k$. Note that for even $k$, the family of modular anyon models $\mathbb{Z}_k^{(n+1/2)}$ is parameterized by $n\in[0,k-1]$, but restricting to the model with $n=0$ yields the central charge corresponding to the $U(1)_k$ conformal field theory. Thus, the full input fusion category contains composite anyons $\{(0,1),(0,\psi),(1,1),(1,\psi),\ldots,(k-1,1),(k-1,\psi)\}$ which obey $\mathbb{Z}_k^{(1/2)}\times \mathbb{Z}_2$ fusion rules. Since $\mathcal{B}$ is Abelian, we still don't have multiple fusion products, but unlike $U(1)_2$, the particles are no longer self-dual, so our lattice is now oriented. The dual of a particle $a=(m,n)$ is given by $\overline{a}=(-m\!\!\mod k,n)$ and the topological spins are given by $\theta_{(m,1)}=e^{2\pi i (m\!\!\mod k)^2/2k}$ and $\theta_{(m,\psi)}=-e^{2\pi i (m\!\!\mod k)^2/2k}$. Analogous to the $U(1)_2$ case, we fix
\al{
    \rho=(0,1)\oplus (1,\psi)
}
consisting of the composite vacuum anyon and the composite charge carrying anyon.

In this model, the Hamiltonian terms of Eqs.~\eqref{eq:Hamiltonian_mass_fusion}-\eqref{eq:Hamiltonian_yang_mills_fusion} become
\al{
    &H_M=\sum_{\bm{v}} P_{\bm{v}}^\psi = \sum_{\bm{v}}(-1)^{\rho_{\bm{v}}}\,, \\
    &\begin{aligned}
        H_K&=\frac{1}{4}\sum_{\bm{e}}(\gamma_{\partial\bm{e}_0}^\po\gamma_{\partial\bm{e}_1}^\po+\overline{\gamma}_{\partial\bm{e}_0}\overline{\gamma}_{\partial\bm{e}_1})\left(O_{\bm{e}}^{\alpha\pd}-O_{\bm{e}}^{\alpha\dagger}\right)  +i(\gamma_{\partial\bm{e}_0}^\po\overline{\gamma}_{\partial\bm{e}_1}-\overline{\gamma}_{\partial\bm{e}_0}\gamma_{\partial\bm{e}_1}^\po)\left(O_{\bm{e}}^{\alpha\pd}+O_{\bm{e}}^{\alpha\dagger}\right) \\
    &=\frac{1}{4}\sum_{\bm{e}}(\mathbb{I}-P_{\partial\bm{e}_0}^\psi)(\mathbb{I}+P_{\partial\bm{e}_1}^\psi)O_{\bm{e}}^{(1,\psi)}+(\mathbb{I}+P_{\partial\bm{e}_0}^\psi)(\mathbb{I}-P_{\partial\bm{e}_1}^\psi)O_{\bm{e}}^{(-1,\psi)} \\
    &=\begin{cases}
        O_{\bm{e}}^{(1,\psi)} & \text{if }\partial\bm{e}_0=(\cdot,\psi)\text{ and }\partial\bm{e}_1=(\cdot,1) \\
        O_{\bm{e}}^{(-1,\psi)} & \text{if }\partial\bm{e}_0=(\cdot,1)\text{ and }\partial\bm{e}_1=(\cdot,\psi) \\
        0 & \text{otherwise}
    \end{cases}\,,
    \end{aligned} \\
    &\!\!\!\!H_{YM}=\frac{g^2}{2}\sum_{\bm{e}}E_{\bm{e}}^2-\frac{1}{a^2g^2}\sum_{\bm{p}}\left(T_{\bm{p}}^{(1,\psi)}+T_{\bm{p}}^{(-1,\psi)}\right)\,,
}
where since $O_{\bm{e}}^{\alpha\pd}\neq O_{\bm{e}}^{\alpha\dagger}$, we have additional constraints on the fermionic hopping. Namely, to hop a fermion ``forward'' we attach a Wilson line $\alpha$ while to hop a fermion ``backward'' we attach a Wilson line $\overline{\alpha}$. For the electric term, the $q$-deformed electric field strength is $\epsilon(a)=[a^2]_q$ with $a\in\mathbb{Z}_k$.

Generalizing the circuit constructions of the $U(1)_2$ $F$ and $R$ symbols to $U(1)_k$ is rather straightforward. Now to encode an anyon in the composite system we require $\lceil\log k\rceil+1$ qubits, with the first $\lceil\log k\rceil$ qubits storing a binary representation of the integer anyon label of the $U(1)_k$ layer, and the last qubit storing the $\{1,\psi\}$ layer. As before, the $F$ symbols of the $\{1,\psi\}$ layer are trivial, so we only need to consider the $F$ symbols of the $U(1)_k=\mathbb{Z}_k^{(1/2)}$ layer, given by
\al{
    [F_{a\oplus_k b\oplus_k c}^{abc}]_{a\oplus_k b,b\oplus_k c}^{\phantom{a}}&=e^{\frac{\pi i}{k}a(b+c-(b\oplus_k c))}\,,
}
where $\oplus_k$ denotes addition modulo $k$. The action of an $F$ move on the encoded states then has the form
\al{
    |a,b;a\oplus_k b\rangle |a\oplus_k b,c;a\oplus_k b\oplus_k c\rangle\mapsto e^{\frac{\pi i}{k}a(b+c-(b\oplus_k c))} |a,b\oplus_k c;a\oplus_k b\oplus_k c\rangle |b,c;b\oplus_k c\rangle\,.
}

\begin{figure}
    \centering
     \begin{subfigure}[t]{0.5\textwidth}
        \centering
            \includegraphics[width=\textwidth]{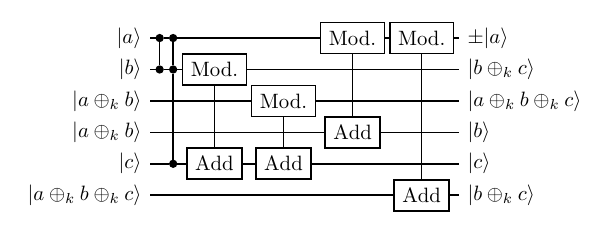}
            \caption{$F$ symbol}
            \label{fig:U(1)_k-Fmove}
     \end{subfigure}\hfill
          \begin{subfigure}[t]{0.5\textwidth}
        \centering
            \includegraphics[width=\textwidth]{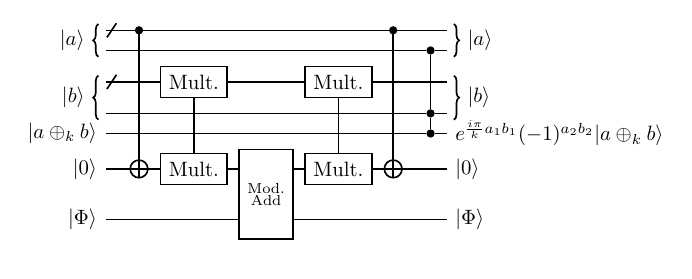}
            \caption{$R$ symbol}
            \label{fig:U(1)_k-Rmove}
     \end{subfigure}
    \caption{\label{fig:U(1)_k-circuits}Explicit circuit implementations for the primitive unitary operations of the $U(1)_k$ LGT Hamiltonian. The (a) $F$ symbol applies a phase depending on the anyons $a$, $b$, and $c$. The $CZ$ gate is controlled on the state $b=k-1$ and applies a Pauli-$Z$ to the last qubit of $a$ which stores the parity of $a$. The $CCZ$ gate is controlled on $b=1$ and $c=k-1$ and similarly applies the Pauli-$Z$ gate. The basis state transformation is then achieved via quantum circuits for in-place modular addition. The (b) $R$ symbol applies a phase depending on the $U(1)_k$ layer by making use of a phase-gradient state $\ket{\Phi}$ and then applies a phase depending on the $\{1,\psi\}$ layer using a $CCZ$ gate.}
\end{figure}

While this transformation holds generally for arbitrary admissible states, we again can drastically simplify the transformation using the fact that $b$ is always given by the generating Wilson line of $U(1)_k$, where in this case since the theory is not self-dual, we have $b=1$ or $b=k-1$. For the case when $b=1$, for all $c\neq k-1$ we have $b+c=b\oplus_k c$, resulting in a trivial phase. When $c=k-1$, we obtain the phase $e^{\pi i a}$. On the other hand, when $b=k-1$, then $b\oplus_k c=c-1$, and we always obtain the phase $e^{\pi i a}$. Since $a$ is an integer, this phase is always just $\pm 1$ depending on the parity of $a$, which can be accounted for using a Pauli-$Z$ gate. Finally, generalizing from addition modulo two, the basis state transformation can be obtained via quantum circuits for in-place addition modulo $k$~\cite{Haner:2020a}. The explicit implementation of this operator is given in Fig.~\ref{fig:U(1)_k-Fmove}.

The overall scaling of this subroutine scales as $O(\log k)$ which is dominated by the modular arithmetic operations~\cite{Vedral:1996a}, since modular arithmetic can be achieved via $O(1)$ applications of binary addition and subtraction subroutines.

Turning now to the $R$ move, we again have contributions from both layers. In the $\{1,\psi\}$ layer, $R^{\psi\psi}_1=-1$ while $R=1$ otherwise. For the $U(1)_k$ layer, the action of the $R$-move on the encoded state is
\al{
     |a,b;a\oplus_k b\rangle\mapsto e^{\frac{\pi i}{k}ab}|a,b;a\oplus_k b\rangle\,.
}
 There are a variety of ways to prepare this phase. The two typical approaches either compute the coefficients directly with quantum circuits for integer and floating-point arithmetic, though some arithmetic operations can be quite costly, or employing a quantum read-only memory (QROM)~\cite{Babbush:2018a}, which loads classically precomputed coefficients into the quantum computer, but can become impractical when there are many unique coefficients. Given the additional structure of these coefficients, we instead implement them via a generalization of the phase-gradient state~\cite{Sanders:2020a}. We define the phase gradient state by
\al{
    |\Phi\rangle =\frac{1}{\sqrt{k}}\sum_{n=0}^{k-1} e^{-2\pi i n/k}|n\rangle\,,
}
where in contrast to the usual phase-gradient state, we let $k$ be an arbitrary integer, not necessarily a power of two. A downside of this approach is that preparing this state becomes significantly more complicated. First one must prepare a uniform superposition of only the target basis states, using for example the \textsc{uniform}$_k$ circuit of Ref.~\cite{Babbush:2018a}. Then one must prepare the unique coefficients on each of these target states, using for example a QROM~\cite{Babbush:2018a}, in which the circuit complexity will scale as $\mathcal{O}(k)$. This is in contrast to the usual phase-gradient state when $k$ is a power of two, which can be prepared with gate cost $\mathcal{O}(\log k\log(\log k/\epsilon))$ when each rotation is implemented to precision $\log( \log k/\epsilon)$~\cite{Bocharov:2015a}. However, provided our phase-gradient state is catalytic, then it need only be prepared once, and its cost should remain negligible in comparison to the remainder of the algorithm\footnote{If one wishes to use to the usual phase-gradient state instead, one must rescale the phase $ab\to \frac{2^{Q-1}}{k}ab$ where $Q$ is the number of qubits in the phase-gradient state. This can be accomplished via a quantum circuit for integer division followed by a bit-shift on the register storing the phase.}. In particular, $|\Phi\rangle$ is an eigenstate of addition modulo $k$, so once we have computed the phase, we can perform modular addition with our phase gradient state since
\al{
    \frac{1}{\sqrt{k}}\sum_{n=0}^{k-1} e^{-2\pi i n/k}|n\oplus_k m\rangle =\frac{1}{\sqrt{k}}\sum_{n=0}^{k-1} e^{-2\pi i (n-m)/k}|n\rangle=e^{2\pi i m/k}|\Phi\rangle \,,
}
and setting $m=ab/2$, we obtain the correct phase on our basis states via phase kickback since $|\Phi\rangle$ is preserved. See Fig.~\ref{fig:U(1)_k-Rmove} for an explicit implementation. In this case, the asymptotic circuit complexity scales as $O(\text{polylog }k)$ resulting from the binary multiplication subroutine~\cite{Parent:2017a}.

%%%%%%%%%%%%%%%%%%%%%%%%%%%%%%%%%%%%%%%%%%%%%%%%%%%%%%%%%%%%%%%%%%%%

\subsubsection{$SU(2)_k$}
\label{sec:SU(2)_k}

%%%%%%%%%%%%%%%%%%%%%%%%%%%%%%%%%%%%%%%%%%%%%%%%%%%%%%%%%%%%%%%%%%%%

As a last example, we consider the non-Abelian $SU(2)_k$ gauge theory, which can be regarded as a $q$-deformation of $SU(2)$ with deformation parameter $q=e^{2\pi i/(k+2)}$. We provide a brief introduction to $q$-deformations of $SU(2)$ along with the relevant input data of the fusion category in Appendix~\ref{app:q-deformed_SU(2)}. The input fusion category is
\al{
    \mathcal{B}=SU(2)_k\boxtimes \{1,\psi\}\,,
}
and following the usual angular momentum representation we label the (self-dual) anyons of $SU(2)_k$ by $j_\ell\in\{\ell/2\}_{\ell=0}^k$. We again fix
\al{
    \rho=(0,1)\oplus (1/2,\psi)
}
consisting of the identity particle and the generating Wilson line of $SU(2)_k$ with the fermion.

In contrast to $U(1)_k$, we now need to include an additional qubit at each dangling edge to store the fermionic occupancy. In particular, this additional qubit allows us to distinguish the two bosonic states $(0,1)|0\rangle$ and $(0,1)|1\rangle$, corresponding to the vacuum and the doubly-occupied fermionic state, respectively. The additional qubit is redundant when hosting a single fermion, since $(1/2,\psi)$ carries the full spin-$\frac{1}{2}$ representation of $SU(2)_k$, so we fix the convention that any dangling edge carrying $(1/2,\psi)$ carries an occupancy in the state $|0\rangle$.

For the kinetic term, the anyons are again self-dual, but the additional qubit breaks this equivalence and introduces an inherent orientation to the hopping term $O_{\bm{e}}^\alpha$ which requires that the fermion statistics are obeyed. For example, a fermion $(1/2,\psi)|0\rangle$ cannot hop to an adjacent doubly-occupied site $(0,1)|1\rangle$, but the reverse process is allowed. Writing down all of the allowed hopping terms, which depend on the fermion number $N_{\bm{v}}\in\{0,1,2\}$ of the two adjacent vertices, we have

The Hamiltonian terms of Eqs.~\eqref{eq:Hamiltonian_mass_fusion}-\eqref{eq:Hamiltonian_yang_mills_fusion}, defined on this larger Hilbert space, are
\al{
    H_M&=\sum_{\bm{v}}P_{\bm{v}}^\psi=\sum_{\bm{v}}(-1)^{\rho_{\bm{v}}}\,, \\
    H_K&=\frac{1}{4}\sum_{\bm{e}}\left(\delta_{N_{\partial\bm{e}_0}+1,N_{\partial\bm{e}_1}}+\delta_{N_{\partial\bm{e}_0}-1,N_{\partial\bm{e}_1}}+\delta_{N_{\partial\bm{e}_0}=1,N_{\partial\bm{e}_1}=1}\right)O_{\bm{e}}^{(1/2,\psi)}\,,\label{eq:SU(2)_kinetic} \\
    H_{YM}&=\frac{g^2}{2}\sum_{\bm{e}}E_{\bm{e}}^2-\frac{1}{a^2g^2}\sum_{\bm{p}}T_{\bm{p}}^{(1/2,\psi)}\,,
}
where the addition associated with the total fermion number in the expression for $H_K$ is performed modulo three and the $q$-deformed electric field strength in $H_{YM}$ is $\epsilon(j)=[j(j+1)]_q$ with $j\in[0,k/2]$.

Unfortunately, while the hopping operator $O_{\bm{e}}^\alpha$ was unitary on the anyonic subspace, once the occupancy qubits are included, the unitarity is broken on the larger space.
In order to reestablish unitarity on this Hilbert space, we can block encode the kinetic term. Since the full Hilbert space of $O_{\bm{e}}^\alpha$ is only 16-dimensional, this procedure can be accomplished straightforwardly using either the sparse-access or LCU block-encoding model~\cite{Low:2019a}. In the sparse-access model, one needs to construct an oracle that determines the locations of the non-zero matrix element values. This is typically accomplished by constructing a quantum circuit which maps input basis states to their corresponding output basis states, but since the outputs are not unique, it will be difficult to find such a general unitary construction. Instead, one must equip the oracle with the known truth table. On the other hand, one can obtain an LCU representation of the kinetic term by searching over all weight-4 Pauli operators and identifying the ones which contribute non-trivially to the Hamiltonian. Doing this, we obtain a Pauli decomposition of $H_K$ containing at most 32 weight-4 Pauli operators such that
\al{
    H_K=\frac{1}{8}\sum_{\bm{e}}\sum_{\ell}\alpha_\ell\sigma_\ell\,,
}
where $\sigma_\ell\in P^{\otimes 4}$ and $\alpha_\ell\in\mathbb{R}^+$ are the coefficients given by the product of $F$ and $R$ symbols in Eqs.~\eqref{eq:hopping_operator_x} and~\eqref{eq:hopping_operator_y}, which we can always take to be positive by absorbing any phases into the Pauli operators.

The LCU oracles can be implemented according to Ref.~\cite{Babbush:2018a}. Namely, using a QROM, the $F$ symbols can be loaded into the quantum computer to prepare an index state in a weighted superposition on five ancilla qubits. Then \textsc{select} is implemented via unary iteration over the 32 index states, and controlled Pauli operations between the index and system registers. Importantly, since the Hilbert space dimension on each edge is fixed, block encoding the hopping operators contributes only $\mathcal{O}(1)$ gate complexity to implementing the kinetic term $H_K$. The remaining Hamiltonian terms can be implemented directly, as in the case of $U(1)_k$, by constructing the explicit $F$ and $R$ symbols as unitary quantum circuits.

We can encode the system in $\lceil\log(k+1)\rceil+1$ qubits, with the first $\lceil\log(k+1)\rceil$ storing the $SU(2)_k$ layer and the last qubit storing the $\{1,\psi\}$ layer. As before, we need only consider the $F$ symbols of the $SU(2)_k$ layer, defined by
\al{
    [F^{j_1j_2j_3}_{j_4}]_{j_5,j_6}=(-1)^{j_1+j_2+j_3+j_4}\sqrt{[2j_5+1]_q[2j_6+1]_q}\left\{\begin{array}{ccc}
        j_1 & j_2 & j_5 \\
        j_3 & j_4 & j_6
    \end{array}\right\}_q\,.
\label{eq:SU(2)_k-coeffs}
}
See Appendix~\ref{app:q-deformed_SU(2)} for additional details regarding these coefficients. Due to the non-Abelian nature of this gauge group, the action on the encoded states generates a superposition over final states,
\al{
    |j_1,j_2;j_5\rangle|j_5,j_3;j_4\rangle\mapsto \sum_{j_6}[F^{j_1j_2j_3}_{j_4}]_{j_5,j_6} |j_1,j_6;j_4\rangle|j_2,j_3;j_6\rangle\,,
}
where the number of terms is determined by the $SU(2)_k$ fusion rules. We again take advantage of the structure of the particular $F$ symbols appearing in our Hamiltonian, allowing us to fix $j_2=1/2$ corresponding to the generating Wilson line of $SU(2)_k$, which further constrains the available $F$ symbols as
\al{
    [F^{j_1 j_2 j_3}_{j_4}]_{j_5, j_6}=[F^{j_1 (1/2) j_3}_{j_4}]_{(j_1 \pm 1/2),(j_3\pm 1/2)}\,.
}

\begin{figure}
    \centering
     \subfloat[$F$ symbol]{\includegraphics[width=0.8\linewidth]{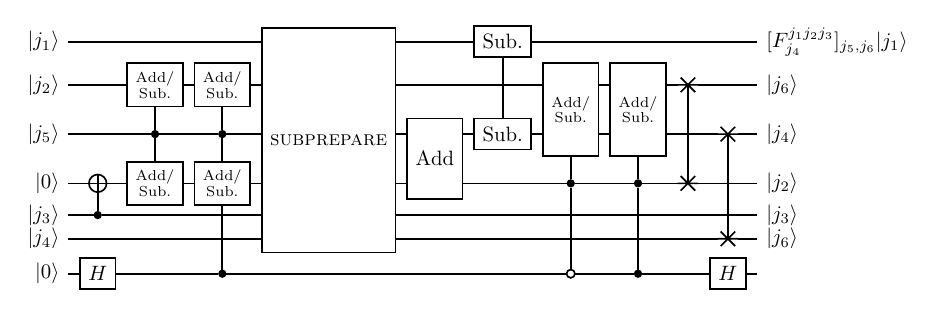}\label{fig:SU(2)_k-Fmove}}
     \newline
     \subfloat[$R$ symbol]{\includegraphics[width=0.8\linewidth]{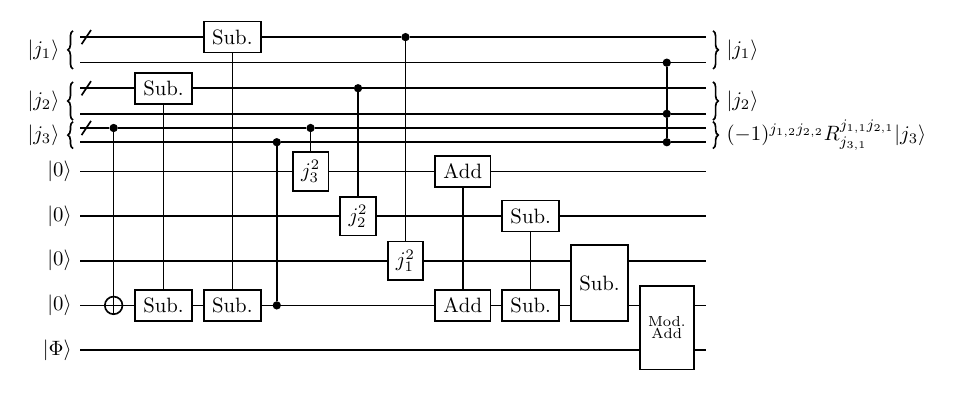}\label{fig:SU(2)_k-Rmove}}
    \caption{Circuits implementing the (a) $F$ symbol for the $SU(2)_k$ LGT. The first three layers prepare the state in Eq.~\eqref{eq:SU(2)_k_initial-state}, and the \textsc{subprepare} oracle of Ref.~\cite{Babbush:2018a} loads and attaches the coefficients to the correct states in the superposition. Binary arithmetic circuits are then used to prepare the state in Eq.~\eqref{eq:SU(2)_k_final-state}, followed by register \textsc{swap} operators to obtain the correct ordering of basis states. (b) $R$ symbol for the $SU(2)_k$ LGT. The catalytic phase-gradient state $|\Phi\rangle$ of Eq.~\eqref{eq:SU(2)_phase-gradient} is used to compute the phase of the $R$ symbol via binary arithmetic on the $SU(2)_k$ layer. The contribution of the fermionic phase is implemented via a $CCZ$ gate on the $\{1,\psi\}$ layer. For simplicity, the uncomputation of the ancilla qubits is not shown, but implicit.}
\end{figure}

Therefore, the action of the $F$ move always generates a superposition over exactly two final states, corresponding to $j_6=j_3\pm 1/2$. This unitary operation corresponds to a generalization of the Clebsch-Gordan transform~\cite{Bacon:2006a}, in which the Clebsch-Gordan coefficients are replaced here with the $q$-deformed Wigner $6j$-symbols. Unfortunately, unlike the Clebsch-Gordan coefficients, there is no known efficient quantum algorithm for computing the (deformed) Wigner $6j$-symbols, meaning they must be classically precomputed and loaded into the quantum computer using the \textsc{subprepare} circuit of Ref.~\cite{Babbush:2018a}. Moreover, the coefficients depend on both the input and output anyons, so the allowed final states $j_6$ must be prepared while preserving the input states $j_1,j_2,\ldots,j_5$ before invoking the \textsc{subprepare} circuit. An explicit realization of this circuit is given in Fig.~\ref{fig:SU(2)_k-Fmove}.

The explicit circuit construction can be understood as follows: The first part of the circuit constructs the state
\al{
    \frac{1}{\sqrt{2}}\left(|j_1,j_2;j_1\pm 1/2\rangle|j_3\pm1/2,j_3;j_4\rangle|0\rangle + |j_1,j_2;j_1\pm 1/2\rangle|j_3\mp 1/2,j_3;j_4\rangle|1\rangle\right)\,,
    \label{eq:SU(2)_k_initial-state}
}
where we have used the simplifications $j_5=j_1\pm1/2$ and $j_6=j_3\pm 1/2$ for the particular $F$ symbols appearing in our Hamiltonian. The \textsc{subprepare} circuit of Ref.~\cite{Babbush:2018a} then attaches the associated coefficient to each state in the superposition, which is a function of each anyon $j_1,\ldots,j_6$. Next, binary arithmetic circuits are used to map $j_5$ to $j_6$ where the corresponding action on $j_5$ can be uniquely determined via controls on both the ancilla state and $j_6$, yielding
\al{
    \begin{aligned}
    &[F^{j_1j_2j_3}_{j_4}]_{j_1\pm 1/2,j_3\pm 1/2}|j_1,j_2,j_3\pm 1/2\rangle|j_3\pm1/2,j_3,j_4\rangle|0\rangle \\
    &\quad+ [F^{j_1j_2j_3}_{j_4}]_{j_1\pm 1/2,j_3\mp 1/2}|j_1,j_2,j_3\mp 1/2\rangle|j_3\mp 1/2,j_3,j_4\rangle|1\rangle\,,
    \end{aligned}
    \label{eq:SU(2)_k_final-state}
}
where the coefficients are defined in Eq.~\eqref{eq:SU(2)_k-coeffs}.

Lastly, the register SWAP gates ensure we end the computation in the correct trivalent encoding. We note that since the ancilla qubit is now entangled to the basis state, it cannot be uncomputed exactly. However, the probability of measuring the ancilla qubit in the zero state is high, so we can continue to use this ancilla qubit for each implementation of an $F$ symbol, and at the end of the computation measure. If the ancilla qubit is zero, then our computation was successful, otherwise, we can repeat the computation.

The dominant cost of this subroutine is the implementation of the \textsc{subprepare} oracle, which scales linearly in the number of unique coefficients. Since $j_2$ is fixed and $j_5$ and $j_6$ have only two possible values, the number of unique coefficients is $O(k^3)$ which determines the cost of the subroutine, as the binary arithmetic will scale at worst $O(\text{log}(k))$.

The fact that this subroutine scales as $O(\text{poly}(k))$ while all remaining subroutines scale at worst as $O(\text{log}(k))$ motivates the need for an efficient quantum circuit computing the (deformed) Wigner $6j$-symbols.

Turning now to the $R$ symbols, we again must accommodate contributions from both layers of the composite anyons. As before, in the $\{1,\psi\}$ layer, the only non-trivial contribution is $R^{\psi\psi}_1=-1$. For the $SU(2)_k$ layer, the action of the $R$-move on the encoded state is
\al{
    |j_1,j_2;j_3\rangle\mapsto (-1)^{j_3-j_1-j_2}q^{\frac{1}{2}(j_3(j_3+1)-j_1(j_1+1)-j_2(j_2+1))}|j_1,j_2;j_3\rangle\,.
}

Since $q$ is a root of unity, we can again use the phase-gradient state techniques from the $U(1)_k$ circuit in Sec.~\ref{sec:U(1)_k}, where now we define
\al{
    |\Phi\rangle=\frac{1}{\sqrt{k+2}}\sum_{n=0}^{k+1}e^{-2\pi i n/(k+2)}\ket{n}
    \label{eq:SU(2)_phase-gradient}
}
as the phase-gradient state. We then prepare the corresponding phase on the phase-gradient state using modular addition and, via phase-kickback, obtain the desired transformation. The explicit circuit implementation is given in Fig.~\ref{fig:SU(2)_k-Rmove}.

There are a few notable differences to point out here in comparison to the $R$ symbols for $U(1)_k$. First, we additionally must compute the parity of $j_3-j_1-j_2$, so we perform this step first, as this phase can be reused since it also appears in phase of $q$. Once $j_3-j_1-j_2$ has been computed on an ancilla, a single $CZ$ gate acting on the last qubit of the ancilla can be used to store the parity. Next, the terms $j_1^2$, $j_2^2$, and $j_3^2$ are computed on ancilla qubits and binary arithmetic operations are performed to obtain the phase of $q$, which is then added to the phase of the $|\Phi\rangle$. Lastly, the contribution from the $\{1,\psi\}$ layer is obtained via a single $CCZ$ gate, as before. We do not show the explicit uncomputation of the ancilla qubits for simplicity, but this step is straightforward.

Again, the asymptotic complexity of this subroutine scales as $O(\text{polylog}(k))$ as a result of the binary multiplication subroutines. However, for the simpler case of squaring a binary number, the constant prefactor can be reduced~\cite{Su:2021a}.

%%%%%%%%%%%%%%%%%%%%%%%%%%%%%%%%%%%%%%%%%%%%%%%%%%%%%%%%%%%%%%%%%%%%
%%%%%%%%%%%%%%%%%%%%%%%%%%%%%%%%%%%%%%%%%%%%%%%%%%%%%%%%%%%%%%%%%%%%

\section{Summary and Outlook}
\label{sec:Conclusion}
%%%%%%%%%%%%%%%%%%%%%%%%%%%%%%%%%%%%%%%%%%%%%%%%%%%%%%%%%%%%%%%%%%%%
%%%%%%%%%%%%%%%%%%%%%%%%%%%%%%%%%%%%%%%%%%%%%%%%%%%%%%%%%%%%%%%%%%%%
In this work, we have constructed explicit KS Hamiltonians coupling fermionic matter to anyonic-regularized gauge fields in order to simulate LGTs on fault-tolerant quantum computers. 
The anyonic-regularized gauge groups we construct here utilize the theory of braided fusion categories and fusion surface models, a generalization of prior work involving $q$-deformations~\cite{Zache:2023b,Hayata:2023a,Hayata:2025a}.
In doing so, we are able to introduce, for the first time, fermionic matter coupled to the regularized gauge field, while retaining systematic bounds on continuum and truncation errors in the model.
We are also able to work with theories beyond (semi-)simple gauge groups, such as $U(N)$ gauge theory.

Additionally, we observe that the Hamiltonian constructions we provide have a simple structure as a linear combination of unitary operators in terms of anyon primitives, $F$ and $R$ symbols.
To this end, we have constructed efficient quantum circuit subroutines for these operators in $U(1)_k$ and $SU(2)_k$ LGTs. 
With the circuit primitives in hand, any state-of-the-art Trotterization or post-Trotter simulation protocol may be applied to obtain resource estimates for state preparation or time evolution of these LGTs. 

Our work also opens up pathways to various prospective research.
An immediate improvement to the work shown here would be focused on the implementation of $F$ symbols, particularly for $SU(2)_k$.
The core routine here would be an implementation of the deformed Wigner-$6j$ coefficients, which to our understanding has no current, efficient implementation besides reading data from quantum memory.
A further step would be the generalization of the construction to $SU(3)_k$ gauge theory, wherein the subtleties of fusion product multiplicities would need to be studied.
Moving beyond $2+1$d is also of interest for realistic simulations of particle physics, wherein $3+1$d formulations in a similar framework may make use of Walker-Wang models~\cite{Walker:2012a}, the higher dimensional generalization of Levin-Wen models.

We believe that the work done here also provides an interesting crossroads for the fields of quantum simulation and fusion surface models and braided fusion categories.
While significant research has already been done in the application of anyons in topological quantum computing~\cite{Simon:2023a, Wang:2010a, RevModPhys.80.1083}, 
the models in use are typically restricted to small-$k$ theories like the Ising or Fibonacci anyons in the context of error correction.
Our work motivates looking at broader classes of braided fusion categories in the context of simulation, and hints at potentially efficient methods for simulation for 2d physics on quantum computers, particularly in the context of condensed matter physics (fractional quantum hall effect and 2d superconductivity).
There is also significant interest in the simulation of the full fusion surface model in the context of topological quantum field theories, rational conformal field theories, and systems with finite non-invertible symmetries~\cite{Inamura:2024a}.

%%%%%%%%%%%%%%%%%%%%%%%%%%%%%%%%%%%%%%%%%%%%%%%%%%%%%%%%%%%%%%%%%%%%
%%%%%%%%%%%%%%%%%%%%%%%%%%%%%%%%%%%%%%%%%%%%%%%%%%%%%%%%%%%%%%%%%%%%
\begin{acknowledgments}
This material is based upon work supported by the U.S. Department of Energy, Office of Science, National Quantum Information Science Research Centers, Quantum Systems Accelerator (Award No. DE-SCL0000121). 

We have benefited from a
number of useful discussions with colleagues, who we
would like to acknowledge (in alphabetical order): Andrew Baczewski, Andrew Landahl, and Jacob Nelson.

Sandia National Laboratories is a multimission laboratory managed and
operated by National Technology and Engineering Solutions of Sandia, LLC., a
wholly owned subsidiary of Honeywell International, Inc., for the U.S.\
Department of Energy's National Nuclear Security Administration under
contract DE-NA-0003525.

This paper describes objective technical results and analysis. Any
subjective views or opinions that might be expressed in the paper do not
necessarily represent the views of the U.S.\ Department of Energy or the
United States Government.
\end{acknowledgments}
%%%%%%%%%%%%%%%%%%%%%%%%%%%%%%%%%%%%%%%%%%%%%%%%%%%%%%%%%%%%%%%%%%%%
%%%%%%%%%%%%%%%%%%%%%%%%%%%%%%%%%%%%%%%%%%%%%%%%%%%%%%%%%%%%%%%%%%%%

%%%%%%%%%%%%%%%%%%%%%%%%%%%%%%%%%%%%%%%%%%%%%%%%%%%%%%%%%%%%%%%%%%%%
%%%%%%%%%%%%%%%%%%%%%%%%%%%%%%%%%%%%%%%%%%%%%%%%%%%%%%%%%%%%%%%%%%%%
%%%%%%%%%%%%%%%%%%%%%%%%%%%%%%%%%%%%%%%%%%%%%%%%%%%%%%%%%%%%%%%%%%%%
%%%%%%%%%%%%%%%%%%%%%%%%%%%%%%%%%%%%%%%%%%%%%%%%%%%%%%%%%%%%%%%%%%%%
\clearpage
\appendix

%%%%%%%%%%%%%%%%%%%%%%%%%%%%%%%%%%%%%%%%%%%%%%%%%%%%%%%%%%%%%%%%%%%%
%%%%%%%%%%%%%%%%%%%%%%%%%%%%%%%%%%%%%%%%%%%%%%%%%%%%%%%%%%%%%%%%%%%%

\section{$SU(2)_k$ Anyon Model}
\label{app:q-deformed_SU(2)}

%%%%%%%%%%%%%%%%%%%%%%%%%%%%%%%%%%%%%%%%%%%%%%%%%%%%%%%%%%%%%%%%%%%%
%%%%%%%%%%%%%%%%%%%%%%%%%%%%%%%%%%%%%%%%%%%%%%%%%%%%%%%%%%%%%%%%%%%%

Here we provide additional details about $SU(2)_k$, which is obtained via a $q$-deformation of the group $SU(2)$. A $q$-deformation is a special case of the broader class of noncommutative differential calculi and involves deforming a Lie algebra via a deformation parameter $q$, which recovers the original algebra in the limit $q\to 1$. See Ref.~\cite{Biedenharn:1995a} for a more general discussion of quantum groups.

Recall that the defining Lie algebra of $SU(2)$, using the usual angular momentum representation, is given by the commutation relations
\al{
    [J^+,J^-]=2J^z\,,\quad [J^z,J^\pm]=\pm J^\pm\,,
}
where $J^{\pm}=J^x\pm J^y$ are the raising and lowering operators. The Casimir operator, corresponding to the total angular momentum squared is
\al{
    J^2=J^+J^- + J^z(J^z-1)\,.
}
The canonical basis is given by the irreps $|j,m\rangle$ with $j\in\{n/2\}_{n=0}^\infty$ and $m\in[-j,j]$, with eigenstates satisfying
\al{
    J^z|j,m\rangle=m|j,m\rangle\,,\quad J^2|j,m\rangle=j(j+1)|j,m\rangle\,.
} 
As it stands, the theory needs to be regularized in order to feasibly encode it on a quantum computer. Rather than directly truncating the maximum angular momentum, we introduce the deformation parameter $q=e^{2\pi i/(k+2)}$, which defines a deformed Lie algebra\footnote{For $SU(N)_k$, the deformation parameter is $q=e^{2\pi i/(k+N)}$.}
\al{\label{eq:deformed_commutation}
    [J^+,J^-]=[2J^z]_q\,,\quad [J^z,J^\pm]=\pm J^\pm\,,
}
with Casimir operator
\al{
    J^2=J^+J^- + [J^z]_q[J^z-1]_q\,,
}
where
\al{
    [n]_q=\frac{q^{n/2}-q^{-n/2}}{q^{1/2}-q^{-1/2}}\,,\qquad [n]_q!=\prod_{m=1}^n [m]_q\,,\quad 
    \label{eq:q-number}
}
are the so-called $q$-numbers, with $[1]_q=1$, $[0]_q=0$, and $[0]_q!=1$. The original algebra is recovered in the limit $q\to1$ since $\lim_{q\to1}[n]_q=n$, as expected. Moreover, since $q$ in this case is a root of unity, $[k+2]_q=[n(k+2)]_q$ for any $n\in\mathbb{Z}$ meaning any state $|j,m\rangle$ is annihilated by the operator $(J^{\pm})^{k+2}$:
\al{
    (J^\pm)^{k+2}|j,m\rangle=0\,.
}
In contrast to the original Lie algebra, or even to the deformed Lie algebra when $q$ is not a root of unity, this reduces the number of irreps even further by hiding the internal degree of freedom $m$, leading to an anyon model with particles indexed solely by their angular momentum $j\leq k/2$.

With the regularized theory in hand, we can now define the input data of the anyon model corresponding to the deformed group $SU(2)_k$, using data from Ref.~\cite{Bonderson:2007a}. The irreps $j_\ell\in\{n/2\}_{n=0}^k$ label the (self-dual) topological excitations, with fusion rules, topological spins, and quantum dimensions given by
\al{
    j_1\times j_2 =\sum_{j_3=|j_1-j_2|}^{\min\{j_1+j_2,k-j_1-j_2\}} j_3\,,\qquad\theta_{j_1}=e^{2\pi i\frac{j_1(j_1+1)}{k+2}}\,,\qquad \text{d}_{j_1}=[2j_1+1]_q\,.
}
The fusion rules must satisfy the additional constraints
\al{
\begin{aligned}
    j_1 &\leq j_2+j_3\,, \\
    j_2 &\leq j_3+j_1\,, \\
    j_3 &\leq j_1+j_2\,, \\
    k &\geq j_1+j_2+j_3\in\mathbb{N}\,.
    \label{eq:fusion_constraints}
\end{aligned}
}

The associated $F$ symbols are
\al{
    [F_{j_4}^{j_1j_2j_3}]_{j_{5}j_{6}}=(-1)^{j_1+j_2+j_3+j_4}\sqrt{[2j_5+1]_q[2j_6+1]_q}\left\{\begin{array}{ccc}
       j_1  & j_2 & j_5 \\
       j_3  & j_4 & j_6
    \end{array}\right\}_q
}
where 
\al{
\begin{aligned}
    \left\{\begin{array}{ccc}
       j_1  & j_2 & j_5 \\
       j_3  & j_4 & j_6
    \end{array}\right\}_q&=\Delta(j_1,j_2,j_5)\Delta(j_5,j_3,j_4)\Delta(j_2,j_3,j_6)\Delta(j_1,j_6,j_4) \\
    &\times \sum_n \bigg[\frac{(-1)^n[n+1]_q!}{[n-j_1-j_2-j_5]_q![n-j_5-j_3-j_4]_q![n-j_2-j_3-j_6]_q![n-j_1-j_6-j_4]_q!} \\
    &\quad\quad\quad\times \frac{1}{[j_1+j_2+j_3+j_4-n]_q![j_1+j_5+j_3+j_6-n]_q![j_2+j_5+j_4+j_6-n]_q!}\bigg]
\end{aligned}
}
are the $q$-deformed Wigner 6$j$-symbols given by the Racah formula and the sum runs over all integers $n_{\max}\leq n\leq n_{\min}$ where
\al{
    n_{\max}&=\max(j_1+j_2+j_5,j_1+j_4+j_6,j_3+j_2+j_6,j_3+j_4+j_5)\,, \\
    n_{\min}&=\min(j_1+j_2+j_3+j_4,j_1+j_3+j_5+j_6,j_2+j_4+j_5+j_6)\,,
}
and we have defined
\al{
    \Delta(j_1,j_2,j_3)=\sqrt{\frac{[-j_1+j_2+j_3]_q![j_1-j_2+j_3]_q![j_1+j_2-j_3]_q!}{[j_1+j_2+j_3+1]_q!}}\,.
}
The $R$ symbols are more simply defined by
\al{
    R_{j_3}^{j_1j_2}=(-1)^{j_3-j_1-j_2}q^{\frac{1}{2}(j_3(j_3+1)-j_1(j_1+1)-j_2(j_2+1))}\,.
}

%%%%%%%%%%%%%%%%%%%%%%%%%%%%%%%%%%%%%%%%%%%%%%%%%%%%%%%%%%%%%%%%%%%%
%%%%%%%%%%%%%%%%%%%%%%%%%%%%%%%%%%%%%%%%%%%%%%%%%%%%%%%%%%%%%%%%%%%%

\section{Convergence of Anyonic-Regularized Hamiltonian}

In order to validate our model, we must ensure that the anyonic-regularized Hamiltonian converges to the KS Hamiltonian in the limit $q\to1$, or equivalently, as $k\to\infty$, in which the braided fusion category $G_k$ converges to the Lie group $G$. Since the anyonic-regularized Hamiltonian is constructed from the product of $F$ and $R$ symbols, we consider the convergence of these quantities in this limit.

\subsection{$U(1)_k$}

Recall that in the $U(1)_k$ LGT, the $F$ and $R$ symbols are defined by
\al{
    [F_{a\oplus_k b\oplus_k c}^{abc}]_{a\oplus_k b,b\oplus_k c}^{\phantom{a}}&=e^{\frac{\pi i}{k}a(b+c-(b\oplus_k c))}\,, \\
    R^{ab}_c&=e^{\frac{\pi i}{k}ab}\,,
}
where $\oplus_k$ denotes addition modulo $k$. Taking products of these operators for a fixed labeling of the state to construct the various Hamiltonian terms will generate an overall phase of the form $e^{\pi i f(a,b,c)/k}$ for some function $f$ (independent of $k$), which in the limit yields
\al{
\lim_{k\to\infty}e^{\frac{\pi i}{k}f(a,b,c)}=1
}
Moreover, the quantum dimension which also appears as a coefficient in Eqs.~\eqref{eq:hopping_operator_x} and~\eqref{eq:hopping_operator_y} is trivial since $\text{d}_1=1$ for the generating Wilson line of $U(1)_k$. Therefore, in the limit $k\to\infty$, the link operators of the full gauge group $U(1)$ are recovered:
\al{
    W_1=\sum_{a\in U(1)_k} |a+1\rangle\langle a|\,, \quad W_1^\dagger=\sum_{a\in U(1)_k} |a-1\rangle\langle a|\,.
}

The convergence of the Casimir operator is also straightforward since we need only check the convergence of the coefficient $[a^2]_q$. Note that the limit $q\to 1$ is equivalent to the limit $k\to\infty$ in the $q$-numbers, and by definition in Eq.~\eqref{eq:q-number}, we have
\al{
    \lim_{q\to 1} [a^2]_q=a^2\,,
}
which recovers the exact Casimir operator of the $U(1)$ LGT. Lastly, we note that since the parity operator $P_{\bm{v}}$ acts only on the fermionic layer, we need not check its convergence.

\subsection{$SU(2)_k$}

Recall that in the $SU(2)_k$ LGT, the $F$ and $R$ symbols are defined by
\al{
    [F_{j_4}^{j_1j_2j_3}]_{j_{5}j_{6}}&=(-1)^{j_1+j_2+j_3+j_4}\sqrt{[2j_5+1]_q[2j_6+1]_q}\left\{\begin{array}{ccc}
       j_1  & j_2 & j_5 \\
       j_3  & j_4 & j_6
    \end{array}\right\}_q\,, \\
    R_{j_3}^{j_1j_2}&=(-1)^{j_3-j_1-j_2}q^{\frac{1}{2}(j_3(j_3+1)-j_1(j_1+1)-j_2(j_2+1))}\,,
}
where $q=e^{2\pi i/(k+2)}$ is the deformation parameter. In the limit $k\to\infty$, or equivalently $q\to 1$, we obtain
\al{
   \lim_{q\to 1} [F_{j_4}^{j_1j_2j_3}]_{j_{5}j_{6}}&=(-1)^{j_1+j_2+j_3+j_4}\sqrt{(2j_5+1)(2j_6+1)}\left\{\begin{array}{ccc}
       j_1  & j_2 & j_5 \\
       j_3  & j_4 & j_6
    \end{array}\right\} \\
    \lim_{q\to 1}R_{j_3}^{j_1j_2}&=(-1)^{j_3-j_1-j_2}\,,
}
for which the usual Wigner-$6j$ symbols are recovered, but there is a persistent phase in the $R$ symbols which must be carefully accounted for.

The convergence of the $SU(2)_k$ pure-gauge theory was demonstrated in Ref.~\cite{Zache:2023b} for the plaquette and Casimir operators. In this case the only $F$ symbols appear in the Hamiltonian, whereas our model additionally contains a single $R$ symbol. However, the particular structure of the $R$ symbol in Eq.~\eqref{eq:plaquette_operator} ensures a trivial contribution since
\al{
    R^{\bar{\alpha}\rho'}_\rho=\begin{cases}
        R_{(1/2,\psi)}^{(1/2,\psi)(0,1)}=1 \\
        R_{(0,1)}^{(1/2,\psi)(1/2,\psi)}=1
    \end{cases}\,,
}
and the contributions from the fermionic layer cancel out any nontrivial contribution from the $SU(2)_k$ layer. Thus, the plaquette operator recovers the usual KS plaquette operator and the Casimir operator converges in the same way as the $U(1)_k$ LGT:
\al{
    \lim_{q\to1}[j(j+1)]_q=j(j+1)\,.
}

\begin{figure}
    \centering
    \includegraphics[width=0.95\linewidth]{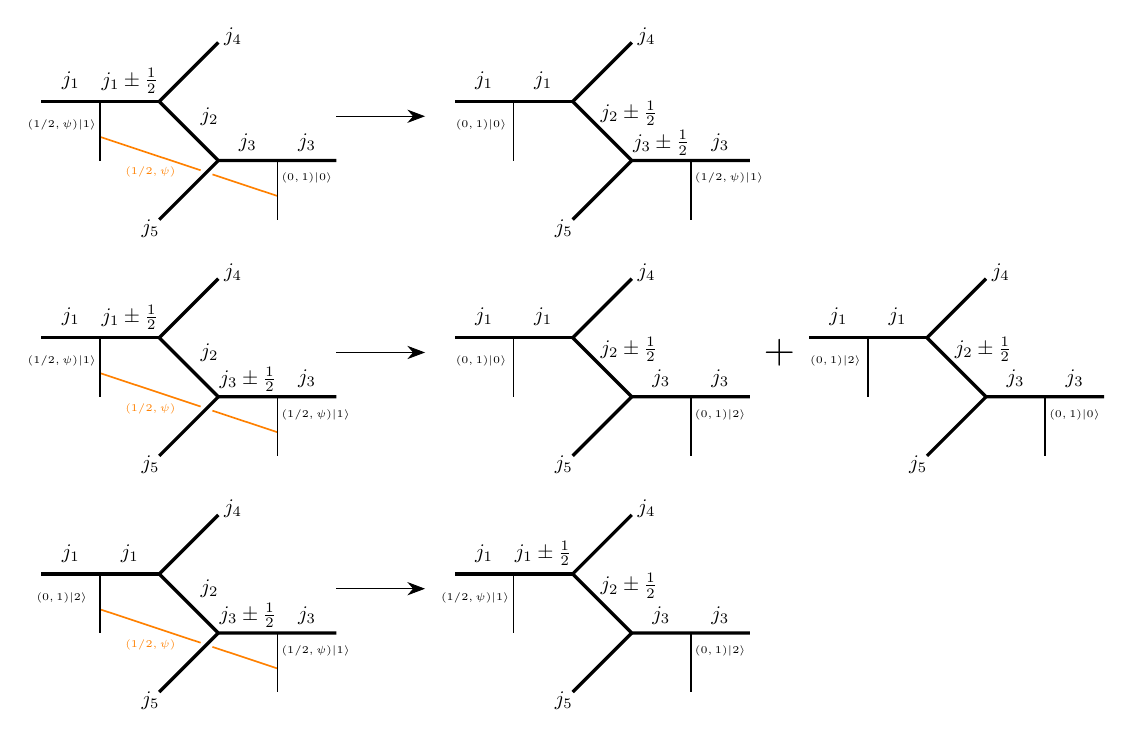}
    \caption{The allowed initial and final states resulting from the application of the forward-hopping term $O_{\bm e}^\alpha$. The reverse-hopping terms will follow analogously. For simplicity, we have not included the specific fermionic layer labels on the non-dangling edges.}
    \label{fig:convergence}
\end{figure}

The only remaining term to consider is the kinetic term. Unfortunately, the $R$ symbols appearing in Eqs.~\eqref{eq:hopping_operator_x} and~\eqref{eq:hopping_operator_y} do not always have a trivial contribution since they depend on non-dangling edges in the lattice which are unconstrained. The application of the kinetic term for $SU(2)_k$ LGTs in Eq.~\eqref{eq:SU(2)_kinetic}, which acts on the enlarged Hilbert space of anyon states and total fermion numbers, ensures we always produce the correct final states. See Fig.~\ref{fig:convergence}.

If we compare our model to that of the spin-network representation of the KS Hamiltonian, it is clear that the nontrivial matrix elements of the KS Hamiltonian will depend only on $j_1$, $j_2$, and $j_3$. However, in our anyonic-regularized model, the coefficients will depend on $j_1,j_2,\ldots,j_5$, with the additional dependencies arising due to the $F$ symbols which are functions of $j_4$ and $j_5$. Thus, our hopping operator will need to be multiplied by a prefactor which depends only upon $j_4$ and $j_5$ in order to correct the amplitudes in our model such that they converge to the same matrix elements as the KS Hamiltonian.

One could compute the hopping amplitudes in the spin-network representation of the KS Hamiltonian and compare to the sequence of $F$ symbols appearing in our anyonic-regularized Hamiltonian to attempt to identify the functional form of these prefactors. If they are simple functions, one could then implement them directly via quantum circuits for arithmetic. However, since there are at most $O(k^2)$ unique coefficients, we can instead classically pre-compute these factors and load them into the computation via a QROM~\cite{Babbush:2018a}. The asymptotic scaling incurred here is still subleading with respect to the cost of computing the $F$ symbols themselves, which scales as $O(k^3)$, so using a QROM will not increase the computational complexity. However, it is interesting to note that since we have included the fermionic layer $\{1,\psi\}$, the amplitudes of our model have the same sign as the usual KS Hamiltonian, and differ in magnitude only.

Thus, after applying these corrective prefactors to the Hamiltonian $H_K$ in Eq.~\eqref{eq:SU(2)_kinetic}, each matrix element of the anyonic-regularized Hamiltonian will converge to the true KS Hamiltonian in the limit $k\to\infty$.

%%%%%%%%%%%%%%%%%%%%%%%%%%%%%%%%%%%%%%%%%%%%%%%%%%%%%%%%%%%%%%%%%%%%
%%%%%%%%%%%%%%%%%%%%%%%%%%%%%%%%%%%%%%%%%%%%%%%%%%%%%%%%%%%%%%%%%%%%

\bibliographystyle{apsrev4-2}
\bibliography{main}

\end{document}